\documentclass[twocolumn,showpacs,preprintnumbers,amsmath,amssymb]{revtex4}

\usepackage[normalem]{ulem}
\usepackage{amsmath}
\usepackage{amssymb}
\usepackage{graphicx}
\usepackage{dcolumn}
\usepackage{bm}
\usepackage{color}

\begin{document}

\title{Classical and quantum dissipative dynamics in Josephson junctions: an Arnold problem, bifurcation and capture into resonance}

\author{Dmitrii Pashin and Arkady M. Satanin}
\affiliation{Department of Theoretical Physics, Lobachevsky State University of Nizhny Novgorod, Nizhny Novgorod 603091, Russia}

\author{Chang Sub Kim}
\email{cskim@jnu.ac.kr (corresponding~author)}
\affiliation{Department of Physics, Chonnam National University, Gwangju 61186, Korea}

\date{\today }

\begin{abstract}
We theoretically study the phase dynamics in Josephson junctions, which maps onto the oscillatory motion of a point-like particle in the washboard potential.
Under appropriate driving and damping conditions, the Josephson phase undergoes intriguing bistable dynamics near a saddle point in the quasienergy landscape.
The bifurcation mechanism plays a critical role in superconducting quantum circuits with relevance to non-demolition measurements such as high-fidelity readout of qubit states.
We address the question ``what is the probability of capture into either basin of attraction'' and answer it concerning both classical and quantum dynamics.
Consequently, we derive the Arnold probability and numerically analyze its implementation of the controlled dynamical switching between two steady states under the various nonequilibrium conditions.
\end{abstract}

\pacs{05.10.-a, 85.25.Cp, 03.65.Yz, 02.50.-r}
\maketitle

\section{Introduction}
\label{Introduction}
Today, superconducting quantum circuits with embedded Josephson junctions can be artificially tailored in hundreds of nanometer sizes, which behave like atoms characterized by well-defined discrete energy levels in the washboard potential \cite{Makhlin2001,You2005,Clarke2008}.
Such a superconducting circuit provides an opportunity for experimentalists to prepare, manipulate, and readout the quantum states at mesoscopic scales.
When two levels, not necessarily energy eigenstates, can be isolated by an appropriate setup, the system forms quantum bits (qubits), which are the primitive building blocks of quantum computers.
Accordingly, superconducting circuits have been one of the intensive research topics from the perspectives of both fundamental interest and realization of quantum computing \cite{Vion2002,Greenberg2002,Cottet2002,Collin2004,Ithier2005,Siddiqi2005,Siddiqi2006,Picot2008,Mallet2009,Nakano2009, Meister2015}.

In this work, we theoretically consider an ac-biased Josephson junction, which serves as a key component of various superconducting circuits.
This paper focuses on the control of dissipative dynamics in the Josephson component, and not on the realization of a qubit or its implication in realistic circuits.
The Josephson junction that we consider enacts a nonlinear oscillator generically composed of an inductor with a gauge-invariant phase variable and an intrinsic capacitor connected in parallel.
To provide a dissipative mechanism, we consider an additional shunt resistor in the circuit in the classical model, and assume a thermal contact with a heat reservoir in the quantum regime.
In addition, we drive the system by applying an alternating current with detuning from the natural frequency.
This electronic component performs as a two-terminal switching device and is a promising bifurcation amplifier with substantial gain, which can be used to measure qubits \cite{Siddiqi2004,Vijay:2009,Kakuyangi2013}.
The ensuing dynamics near a bifurcation point is very sensitive to perturbation; for instance, a weak interaction with a qubit will subserve the perturbation.
By assuming the feasibility of controlling the initial states of the nonlinear system, we monitor the relaxation dynamics numerically in great detail.
Consequently, we are able to describe quantitatively how a bifurcation develops temporally in the system, which is an essential feature for the switching function, both in the classical and quantum regimes.
The classical description prevails when the thermal energy exceeds the energy scale defined by the characteristic frequency of the Josephson oscillator, providing an insightful physical picture of the nonlinear dynamics in the complex energy landscape.
However, at low temperatures, the thermal energy is negligible compared to the oscillator energy-level spacing, and therefore a crossover to quantum regime is expected.

The primary goal of this paper is to address the question regarding the probability of the Josephson oscillator being captured onto one of the stable steady states in the classical and quantum regimes.
This problem was studied earlier in a charged particle motion under an electromagnetic field \cite{Lifshitz1962}.
It was rigorously framed into mathematical mechanics by V. I. Arnold \cite{Arnold1963} and is known as an Arnold problem in nonlinear dynamics \cite{Neishtadt1991,Haberman1995,Arnold:Book2006}.
The indeterministic nature of the branching on damping in the phase Josephson junctions was reported experimentally by others, where the Josephson junctions were operated in the classical regime \cite{Goldobin2013,Goldobin2016}.
Moreover, it was revisited recently in simple quantum dynamics by the present authors \cite{Pashin2017}.
The switching between two coexisting stable states in the classical Josephson oscillator has previously drawn much research attention.
However, to our knowledge, the consideration of the probability formulation in quantum dynamics problems is rare.
Here, we explore the classical and quantum Arnold problems in a coherent treatment in a single report.

We describe the Josephson junction by adopting the conventional one-particle Hamiltonian of the macroscopic phase, where the terms specifying charging energy and power input play the roles of kinetic and potential energies, respectively.
We then conceive of the behavior of the Josephson phase as a fictitious particle, which we name the `Josephson particle'.
We approximate the potential energy up through quartic terms, thus framing our problem in terms of a Duffing oscillator in nonlinear dynamics \cite{Nayfeh1979}.
The external current adds a time-dependent term to the potential energy, which renders the effective Hamiltonian nonautonomous.
The Duffing oscillators with quartic nonlinearity, driven by a time-dependent periodic force, have been studied widely in nonlinear dynamics problems.
For instance, a series of applications of the Duffing equation have appeared \cite{Dykman1979,Dykman1988,Dykman2007,Andre2012}, where it has been reported that fluctuations provide a mechanism to overcome the dynamical barrier.
Consequently, they induce switching between the stable solutions, accompanied by dissipation of the oscillator states.
We also focus our attention on other works that are devoted to the tunneling problem between two stable attractors
\cite{Sazonov1976,Dmitriev1986,Wielinga1993,Peano2004,Marthaler2007}.
However, unlike our contribution to the quantum formulation, most of the previous works were based on the semiclassical approximation.
In addition to the cited above works, we also refer to the review articles in which these and related subjects are described \cite{Dykman1984, Grifoni1998, Dykman2012}.
We emphasize that our study reveals the temporal development of the Arnold probability, continually covering time scale before and after the first bifurcation toward a steady state, which gives a new insight into the bifurcation dynamics in classical Josephson junctions.
Moreover, in the quantum case, the bifurcation dynamics in terms of the Duffing oscillator has not been considered previously; therefore our investigation provides a largely new outcome.

Although our work aims at a theoretical manifestation of the Arnold dynamics, it seems worthwhile to note its experimental relevance to an operating Josephson junction in the quantum domain \cite{Likharev1983}.
Recently, there was speculation among researchers on the existence of discrete levels in and quantum tunneling out of a Josephson washboard potential \cite{Blackburn2017,Blackburn2016}.
The authors claim that the selected experimental data of the switching current distribution at low temperatures can be explained classically without resorting to the quantum escape rate \cite{Voss1981,Martinis1985,Yu2010}; accordingly, even below the crossover temperature, a Josephson junction may not be fully quantum mechanical.
We consider the raised question fundamental and challenging but still disputable, requiring further investigation for settlement.
We notice that the Josephson junctions they analyzed is typically in the micrometer range, and the quantum crossover is driven by lowering the temperature.
We believe that apart from the low-temperature condition, there is another mechanism that plays a role in the quantum crossover, which is the length scale.
Specifically, we consider the nanometer-sized Josephson junctions, which set a mesoscopic quantum domain \cite{Barone2010}, where it is expected that the Josephson phase states are quantized.
The modern nano-technique allows researchers to fabricate such mesoscopic Josephson junctions with a high Q-factor and nonlinearity in the laboratories.
Thus, the physical regime that suits our purposes, where the level spacing is far bigger than the level broadenings from both temperature and environmental noise, can be reached so that our prediction of the Arnold formula may be realized.

Finally, we mention the relevance of the studied dynamic Arnold bifurcation to different physical systems.
Recent nanomechanical resonators are other systems where our predicting novel nonlinear effects may be manifested because these systems are characterized by extremely high frequency with relatively weak dissipation and a high Q-factor \cite{Greenberg2012, Rips2014,Chaste2011,Laird2012}.
The theoretical description developed in this work would be beneficial in exploring the nonlinear dissipative dynamics of a quantum state in such a mesoscopic system.

The rest of this paper is organized as follows.
In Sec.~\ref{Arnold-dynamics}, we address the Arnold question in a simple model to introduce the basic concept.
In Sec.~\ref{Classical-dynamics}, we study the classical dynamics of the Josephson particle in phase space and establish the Arnold probability.
Then, we treat the problem quantum mechanically in Sec.~\ref{Quantum-dynamics}:
In Sec.~\ref{single-particle spectrum} we solve the quasi-stationary eigenvalue problem, and in the subsequent Sec.~\ref{DMformulation}, by employing density matrix formalism we investigate the relaxation dynamics of the pumped system in a heat bath and derive the quantum analog of the Arnold formula.
Finally, in Sec.~\ref{Conclusion}, we provide the summary and conclusion.

\section{The Arnold problem}
\label{Arnold-dynamics}
Here, we recapitulate the original Arnold problem, which considers a classical particle with mass $m$ in a static double-well potential \cite{Arnold1963}.
The motion of the particle is governed by the equations of motion in phase space,
\begin{eqnarray*}
\dot{x} =\frac{p}{m}\quad{\rm and}\quad
\dot{p} =-\frac{\partial V(x)}{\partial x}- \gamma p,
\end{eqnarray*}
where $V(x)$ represents the potential energy and $\gamma$ is a damping coefficient.

\begin{figure}[h!]
\begin{center}
\includegraphics[width=0.35\textwidth]{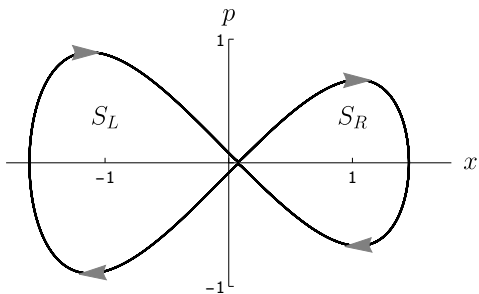}
\caption{Separatrix of a particle in a double-well potential in phase space, exhibiting two competing stable centers separated by an unstable equilibrium point; $S_L$ and $S_R$ denote the areas enclosed by the left and right lobes, respectively, and the arrow shows the direction of the trajectory.}
\label{Fig1:separatrix1}
\end{center}
\end{figure}

The \textit{separatrix} is a phase portrait of the particle with energy matching the top value ($\equiv V_0$) of the central barrier in the double well, which is an ideal trajectory in phase space within the limit of vanishing dissipation, $\gamma\rightarrow 0$.
Such a separatrix is depicted in Fig.~\ref{Fig1:separatrix1}, where the crossing point corresponds to the unstable equilibrium point of the potential energy.
The separatrix defines the boundary in phase space, separating two distinct modes in particle motion under damping: 1) the states outside the separatrix will tend to cross a point on one of the two branches of the separatrix curve in time; 2) the states inside the separatrix will relax to the stable equilibrium point in the corresponding basin of attraction.

Within the small damping limit, by applying the work-energy theorem, one can calculate the energy changed over a full cycle of the trajectory, which is approximately close to the separatrix, to the linear order in $\gamma$.
The outcome is
\[
\Delta H = -\gamma \int p{\dot x} dt \rightarrow -\gamma\oint pdx = -\gamma (S_L+S_R).
\]
If the particle initiates its motion from a phase point randomly chosen outside the separatrix, it will eventually cross the separatrix after a long time by dissipation to stochastically enter one of the lobes.
It is well known that the probability of capture into each basin of attraction is proportional to the change in the Hamiltonian during one cycle of the corresponding homoclinic trajectory \cite{Haberman1995}.
Accordingly, the probability $P_\alpha$ of the particle falling onto either the left (L) or right (R) equilibrium states would be proportional to the bounded area of each lobe, $S_L$ and $S_R$, respectively.
We call the resulting probability $P_\alpha$ as the Arnold formula.
It is expressed as
\begin{equation}\label{ArnoldFormula}
P_\alpha = \frac{S_\alpha}{S_L+S_R}, \quad \alpha=L,\ R,
\end{equation}
which is evidently independent of the damping strength.

\section{Single Josephson junction: Classical dynamics}
\label{Classical-dynamics}
Here, we formulate the Arnold problem in classical Josephson junctions and investigate the dissipative dynamics of the system before and after bifurcation toward a steady state. Most previous works consider the static switching behavior near a bifurcation point without studying the full development before and after bifurcation toward a steady state.

A Josephson junction is macroscopically described by the supercurrent of electron pairs and the voltage across the junction, where the current is expressed as $I=I_c \sin\varphi$ with $\varphi$ representing the relative phase of the condensed states between two superconducting sides and $I_c$ representing the critical current, and the voltage is expressed as $V=\bar\Phi_0 \dot\varphi$, where $\bar\Phi_0=\hbar/(2e)$ is the reduced magnetic-flux quantum.
Accordingly, the classical Hamiltonian for the Josephson junction in the presence of the external current $I_{ex}$ may be written as
\begin{equation}
\label{cHam}
H = \frac{1}{2}C(\bar\Phi_0{\dot\varphi})^2 + E_J(1-\cos\varphi) - \bar\Phi_0 I_{ex}\varphi.
\end{equation}
The first term on the right hand side (RHS) of Eq.~(\ref{cHam}) is the charging energy $\frac{1}{2}CV^2$ in the junction, where $C$ is the intrinsic capacitance of the Josephson junction.
The second term is a potential energy associated with the power input, $VI\sim \dot\varphi\sin\varphi$, where $E_J=I_c\bar\Phi_0$ represents the Josephson energy.
The last term describes time-dependent external control by the driving current, $I_{ex}(t)=I_0\cos(\omega t)$.
This Hamiltonian can be viewed as a function of two canonically conjugate variables, $\varphi$ and $p_\varphi=C\bar\Phi_0^2\dot\varphi$, i.e., $H=H(\varphi,p_\varphi)$.
Then, the standard Hamiltonian formulation yields
\begin{eqnarray*}
\dot \varphi &=& \frac{\partial H}{\partial p_\varphi}=\frac{1}{C\bar\Phi_0^2}p_\varphi, \\
{\dot p}_\varphi &=& - \frac{\partial H}{\partial \varphi} = - E_J\sin\varphi +\bar\Phi_0I_{ex},
\end{eqnarray*}
which generate the Newtonian equation of motion for the phase variable, $C\bar\Phi_0^2\ddot{\varphi} + E_J \sin\varphi = \bar\Phi_0I_{ex}$.
In addition, to account for dissipation, we insert the Ohmic current given by $V/R$, with $R$ being the intrinsic resistance, in the preceding equation to generalize it as
\begin{equation}
\label{cleq2}
\ddot{\varphi}+\frac{1}{RC}\dot{\varphi}+\frac{E_J}{C\bar\Phi_0^2} \sin\varphi=\frac{1}{\bar\Phi_0 C}I_{ex}.
\end{equation}
Then, we define the various coefficients as
\begin{equation}
\label{parameters1}
\gamma\equiv\frac{1}{RC},\quad \omega_0\equiv\sqrt{\frac{E_J}{C\bar\Phi_0^2}},\quad \omega_e\equiv\sqrt{\frac{I_0}{\bar\Phi_0C}}
\end{equation}
and further introduce the various dimensionless variables as
\[\tau\equiv\omega_0 t,\quad \bar\gamma\equiv\frac{\gamma}{\omega_0},\quad {\bar I}_0\equiv\frac{ \omega_e^2}{\omega_0^2},\quad {\rm and}\quad \bar \omega\equiv\frac{\omega}{\omega_0}.\]
With these arrangements, the equation of motion, Eq.~(\ref{cleq2}), is cast into the dimensionless form
\begin{equation*}
\label{cleq3}
\frac{d^2\varphi}{d\tau^2} + {\bar\gamma}\frac{d\varphi}{d\tau} + \sin\varphi = {\bar I}_0 \cos(\bar \omega \tau),
\end{equation*}
which represents a nonlinear, damped harmonic oscillator with pumping.
For a comprehensive analysis, we shall work in the regime where the relative phase is small across the junction so that we can approximate $\sin\varphi$ up to the cubic term.
Then, the equation of motion to be analyzed becomes a Duffing equation expressed as
\begin{equation}
\label{cleq4}
\ddot{\varphi} + {\bar\gamma}\dot{\varphi} + \varphi-\frac{1}{6}\varphi^3 = {\bar I}_0 \cos(\bar \omega \tau),
\end{equation}
where $\cdot$ is understood to be the time-derivative with respect to the dimensionless time $\tau$.
Note that Eq.~(\ref{cleq4}) takes a generic form which includes the conservative, dissipative, and external forces.

Here, we find it useful to parameterize the phase variable $\varphi$ in terms of the \textit{in-phase} $(q)$ and \textit{quadrature-phase} $(p)$ components with respect to the driving oscillation \cite{Siddiqi2004,Bogoliubov1962}.
To this end, we convert the dynamical variables $(\varphi,\dot\varphi$) into $(q,p)$ as
\begin{eqnarray}
&&\varphi = q \cos(\bar \omega \tau) + p \sin(\bar \omega \tau), \label{trans1} \\
&&\bar \omega^{-1}\dot{\varphi} = -q \sin(\bar \omega \tau) + p \cos(\bar \omega \tau), \label{trans2}
\end{eqnarray}
wherein the second equation gives rise to a constraint $\dot{q} \cos(\bar \omega\tau)+\dot{p} \sin(\bar \omega\tau)\equiv 0$.
Consequently, after some manipulation, one can obtain the equations of motion for the new variables.
We then take the time-average of the resulting equations over the half-period, $\pi/\bar \omega$, of the external force to smooth out the fast oscillatory time-dependence.
The outcome of this slowly varying amplitude approximation (SVAA) is given as
\begin{eqnarray}
\dot{p} &=& -\alpha q+\beta q (q^2+p^2)+f-\frac{{\bar\gamma}}{2}p, \label{effeq1}\\
\dot{q}& =& \alpha p-\beta p (q^2+p^2)-\frac{{\bar\gamma}}{2}q, \label{effeq2}
\end{eqnarray}
where the coefficients are defined as
\[
\alpha\equiv\frac{1-\bar \omega^2}{2 \bar \omega},\quad \beta\equiv\frac{1}{16\bar \omega}, \quad{\rm and}\quad f\equiv\frac{{\bar I}_0}{2 \bar \omega}.
\]
Note that there appears a non-conventional damping term, $-\frac{1}{2}{\bar\gamma} q$, in Eq.~(\ref{effeq2}).

Equations~(\ref{effeq1}) and (\ref{effeq2}) constitute the coarse-grained equations of motion of the classical variables describing the Josephson dynamics.
For further analysis, we find it useful to construct the effective Hamiltonian function that generates the conservative dynamics in Eqs.~(\ref{effeq1}) and (\ref{effeq2}).
By inspection, we have obtained the effective Hamiltonian as
\begin{equation}\label{clHam2}
\tilde H(q,p)=\frac{\alpha}{2}(q^2+p^2)-\frac{\beta}{4}(q^2+p^2)^2-fq,
\end{equation}
where $\tilde H$ is normalized with respect to the reference energy $E_{nor}\equiv\omega\omega_0C\bar\Phi_0^2$, which gives the same energy scale as the original Hamiltonian expressed in Eq.~(\ref{cHam}).
\begin{figure}[h!]
\begin{center}
\includegraphics[width=0.47\textwidth]{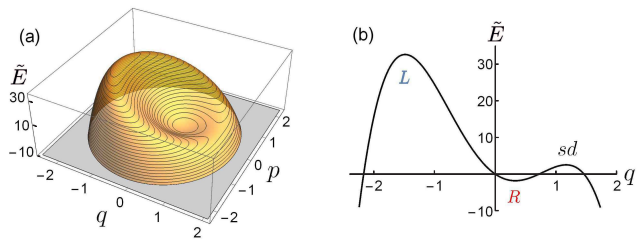}
\caption{Illustration of the quasienergy landscape: (a) Quasienergy surface ${\tilde E}(q,p)$, (b) Quasienergy contour ${\tilde E}(q,p_{sd})$ with $p_{sd}=0$, where the energies are in rescaled energy units of $\hbar\omega_0$. Three fixed points correspond to the large-amplitude state ($L$-state), small-amplitude state ($R$-state), and saddle point ($sd$).}
\label{Fig2:Esurface}
\end{center}
\end{figure}
Unlike the usual Hamiltonian, this function cannot be separated into kinetic energy and potential energy.
However, one may still define the quasienergy of the system as an instant value of the Hamiltonian function, i.e., given $q=q(t)$ and $p=p(t)$,
\[{\tilde E} \equiv {\tilde H}(q,p).\]
As a result of the SVAA, the original oscillatory time-dependence has disappeared in Eq.~(\ref{clHam2}) so that the Hamiltonian becomes approximately \textit{autonomous}.
We present below the numerical data for the energies in rescaled units of $\hbar\omega_0$ for later analysis.
In addition, for reflecting the cubic-order expansion of the potential force $\sim \sin\varphi$, we consider the quasienergy only in the range of ${\tilde E}\ge -10$.

In Fig.~\ref{Fig2:Esurface}, we illustrate the typical quasienergy surfaces described by Eq.~(\ref{clHam2}), where we have assumed a driving force as $f=0.04$ and the numerical values for $\alpha$ and $\beta$ as
\[ \alpha=1.30\times 10^{-1}\quad {\rm and}\quad\beta=7.11\times 10^{-2},\]
which have been estimated using the physical parameters in an experiment \cite{Siddiqi2004}.
We shall use these values throughout the following calculation.
The quasienergy landscape appears intriguing, as shown in Fig.~\ref{Fig2:Esurface}(a).
The steady-state condition generates three fixed points, all of which happen to be at $p=0$; two of them are stable and the other is a saddle point.
One of the two stable extrema appears on the top surface, $(-1.49,0)\equiv(q_L,p_L)$, with quasienergy ${\tilde E}_L$, and the other at the bottom of the well, $(0.32,0)\equiv(q_{R},p_{R})$, with quasienergy ${\tilde E}_{R}$.
We have performed linear stability analysis to confirm that both points are \textit{centers}.
The saddle point appears at $(q_{sd},p_{sd})=(1.16,0)$ with quasienergy ${\tilde E}_{sd}$.
These features are better viewed in Fig.~\ref{Fig2:Esurface}(b), which depicts the quasienergy contour on the $(q,0)$-plane.
For reference, we provide below the numerical values of ${\tilde E}_{L}$, ${\tilde E}_{sd}$, and ${\tilde E}_{R}$ determined from the adapted parameters.
\[{\tilde E}_{L}=33.1,\ {\tilde E}_{sd}=2.60, \ {\rm and}\ {\tilde E}_{R}=-1.79\]
In the following discussion, we shall call the phase point at $(q_{L},p_{L})$ as the $L$-point and the one at $(q_{R},p_{R})$ as the $R$-point, where $L$ and $R$ symbolize `left' and `right' of $q=0$ on the $q$ axis, respectively.

\begin{figure}[h!]
\begin{center}
\includegraphics[width=0.3\textwidth]{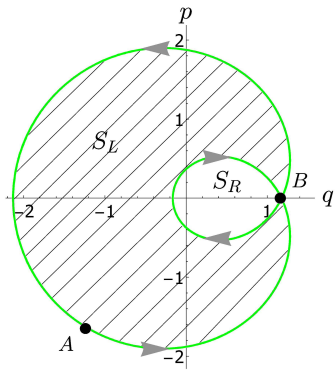}
\caption{The separatrix in green dividing the mechanical states in phase space, on which the particle ideally travels counterclockwise along the outer loop, which is a holonomic orbit, in the vanishing dissipation limit, whereas it travels along the inner holonomic orbit clockwise before completing a full cycle. The shaded area is denoted by $S_L$ and the enclosed area in the inner lobe by $S_R$. The crossing point indicated by $B$ corresponds to the saddle point toward which the two homoclinic orbits approach. Note that the black dot marked by $A$ has been inserted for later use.}
\label{Fig3:separatrix2}
\end{center}
\end{figure}
Then, we view Eqs.~(\ref{effeq1}) and (\ref{effeq2}) as generalized Hamiltonian dynamics generated by the effective Hamiltonian, Eq.~(\ref{clHam2}), with the additional dissipative terms
\begin{eqnarray}
\dot{p}&=& -\frac{\partial \tilde H}{\partial q} - \frac{{\bar\gamma}}{2} p, \label{Hameq1}\\
\dot{q} &=& \frac{\partial \tilde H}{\partial p} - \frac{{\bar\gamma}}{2} q. \label{Hameq2}
\end{eqnarray}
Note that the dynamical description given by Eqs.~(\ref{Hameq1}) and (\ref{Hameq2}) contains not only a generalized force but also a generalized velocity, exhibiting an extended dynamics in the generic form \cite{Kim:2015}.

As in the original Arnold problem, the trajectories generated by the Josephson dynamics may be classified by a \textit{separatrix}, which is the unperturbed phase-trajectory to which the particle with the saddle point energy ${\tilde E}_{sd}$ tends ideally in the limit of vanishing dissipation, ${\bar\gamma}\rightarrow 0$.
In Fig.~\ref{Fig3:separatrix2}, we depict a separatrix resulting from the parameters $\alpha$ and $\beta$ used in Fig.~\ref{Fig2:Esurface} for the same $f=0.04$, which comprises two homoclinic orbits, a homoclinic orbit being the phase trajectory joining a saddle point to itself.
Divided by the separatrix, the energies of the states are categorized into three classes: 1) the phase points inside the inner lobe are limited within the quasienergy window ${\tilde E}_{R}<{\tilde E}<{\tilde E}_{sd}$; 2) the states confined between the outer homoclinic and the inner homoclinic (shaded area) possess energies greater than the saddle-point energy, i.e., ${\tilde E}>{\tilde E}_{sd}$; 3) the states outside the larger homoclinic possess energies less than the saddle-point energy i.e., ${\tilde E}<{\tilde E}_{sd}$.

\begin{figure}[h!]
\begin{center}
\includegraphics[width=0.45\textwidth]{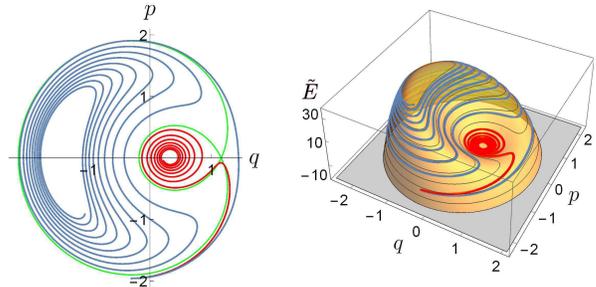}
\caption{[Color online] Two dissimilar phase-trajectories from the same initial condition at $(q_0,p_0)=(-0.3,-1.96)$ located outside the large separatrix in Fig.~\ref{Fig3:separatrix2}; the blue and red colored trajectories are the outcomes of ${\bar\gamma}= 0.003$ and $0.007$, respectively, for a given $f=0.04$. For illustration purposes, the separatrix has been inserted as a green curve. The chosen initial quasienergy ${\tilde E}_0=-1.84$ lies below ${\tilde E}_{sd}$, which is actually below ${\tilde E}_{R}$ in this particular case. The same trajectories are also drawn on the quasienergy surface ${\tilde E}={\tilde H}(q,p)$.}
\label{Fig4:trajectories}
\end{center}
\end{figure}
We performed numerical experiments to monitor the ensuing trajectories by solving Eqs.~(\ref{Hameq1}) and (\ref{Hameq2}).
To compute the dynamics, an arbitrary initial state has been chosen in each of the three distinct regions in phase space for two different damping strengths.
The outcome is as follows:
1) when an initial state is chosen inside the inner lobe, the trajectory falls on the $R$-focus regardless of the damping magnitude;
2) when the initial state is located inside the shaded area in Fig.~\ref{Fig3:separatrix2}, the trajectory falls on either the $L$-focus or the $R$-focus depending on the damping strength;
3) when we choose an  initial state outside the large separatrix, there is a finite probability of the resulting trajectory falling on either focus for a finite damping, as in case 2).
In Fig.~\ref{Fig4:trajectories}, we depict the trajectories for the case 3).

Having learned that the dynamics generated from the initially chosen states placed outside the separatrix in phase space manifests a bistable behavior within the small damping limit, it is of interest to determine the respective probabilities of the resulting trajectories to relax stochastically on either the $L$- or the $R$- point.
This constitutes the Arnold problem.
To study the Arnold question quantitatively, we consider the change in the quasienergy along a segment of the separatrix for motion with a small damping ($\bar\gamma\ll 1$), which can be calculated as
\begin{eqnarray}\label{work-energy2}
\Delta {\tilde H} &=& \int \left( \frac{\partial {\tilde H}}{\partial p}\dot p + \frac{\partial {\tilde H}}{\partial q}\dot q\right)d\tau \nonumber \\
&=& \int \left\{ (\dot q + \frac{{\bar\gamma}}{2}q)\dot p + (-\dot p -\frac{{\bar\gamma}}{2}p)\dot q  \right\}d\tau \nonumber\\
&=& \frac{{\bar\gamma}}{2}\int (q\dot p - p\dot q)d\tau.
\end{eqnarray}
Although the phase-trajectories are not exactly periodic for the considered dissipative dynamics, we may assume that within the limit of vanishing damping, the particle returns to an initial phase point after a cycle within an error of $O(\bar\gamma)$.
Accordingly, the quasienergy change of the particle along the outer closed loop of the separatrix, which is a homoclinic orbit, is evaluated as
\begin{widetext}
\begin{eqnarray}\label{outergain}
\Delta {\tilde H}_{out} &=& \frac{{\bar\gamma}}{2}\oint_{out}(qdp-pdq)
= \frac{{\bar\gamma}}{2} \left\{ (S_L+S_R) - \left(- (S_L+S_R)\right)\right\}\nonumber\\
&=&{\bar\gamma} (S_L+S_R),
\end{eqnarray}
\end{widetext}
where $S_R$ is the phase-space area enclosed in the inner lobe, and $S_L$ is the area enclosed in the outer lobe subtracted by $S_R$, which constitutes the shaded area.
Similarly, the quasienergy change along the inner closed loop, which is again a homoclinic orbit, is calculated to be
\begin{equation}\label{innnerloss}
\Delta {\tilde H}_{in} = \frac{{\bar\gamma}}{2} ( -S_R - S_R) = -{\bar\gamma} S_R.
\end{equation}
Thus, the quasienergy increases when the particle evolves around the outer loop in the counterclockwise manner, whereas it loses energy along the clockwise inner loop of the separatrix.

\begin{figure}[h!]
\begin{center}
\includegraphics[width=0.3\textwidth]{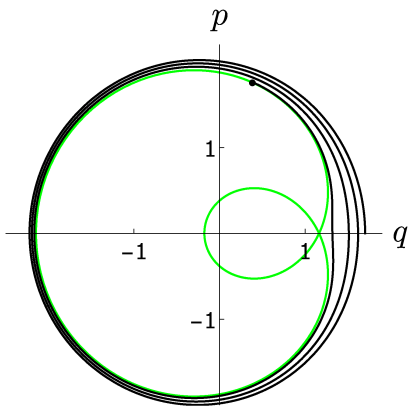}
\caption{A trajectory in state space, which initiates from the state $(q_0,p_0)=(1.70,0)$ with quasienergy ${\tilde E}_0=-7.9<{\tilde E}_{sd}$ [this initial quasienergy happens to be below ${\tilde E}_{R}$], $f=0.04$, and ${\bar\gamma}=0.001$;
we have inserted the separatrix to show that the trajectory initially swirls around the outer separatrix in the counter-clockwise manner and will eventually cross the separatrix and enter inside. Note that we depict the entering point by a dot on the green-colored separatrix.}
\label{Fig5:Entering}
\end{center}
\end{figure}
\begin{figure}[h!]
\begin{center}
\includegraphics[width=0.45\textwidth]{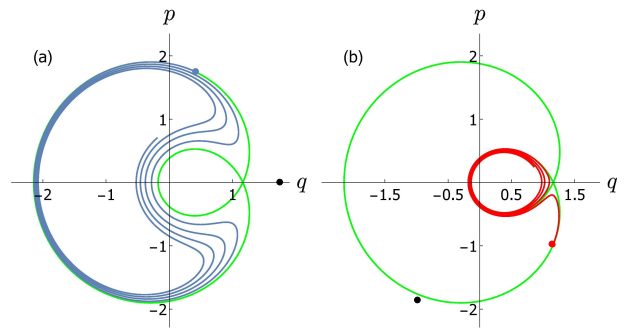}
\caption{[Color online] Two distinctive ensuing trajectories immediately after `entering' taking places onto the separatrix: (a) From initial state $(q_0,p_0)=(1.70,0)$; (b) From initial state $(q_0,p_0)=(-1,-1.87)$; where all other parameters are the same as in Fig.~\ref{Fig5:Entering}. The initial states are marked by the black dots in both cases and the entering points are marked by the blue and red dots, respectively, on the green-colored separatrix.}
\label{Fig6:Afterentering}
\end{center}
\end{figure}
Next, we choose an initial state $(q_0,p_0)$ outside the external contour of the separatrix in Fig.~\ref{Fig3:separatrix2}, whose corresponding quasienergy ${\tilde E}_0$ is below ${\tilde E}_{sd}$, and simulate its motion for a small damping $\bar\gamma$.
Fig.~\ref{Fig5:Entering} illustrates the initial trajectory; it is seen that the trajectory swirls counterclockwise about the outer region of the separatrix.
The particle keeps evolving while gradually reducing its radius of curvature with time.
In this process, the quasienergy of the particle progressively augments and must reach the value ${\tilde E}_{sd}$ at a certain time.
At this moment, we say that the particle `enters' the outer separatrix, and have indicated such an entering point by the notation $A$ on the separatrix loop in Fig.~\ref{Fig3:separatrix2}.
We observe that upon entering, the ensuing trajectory follows the outer separatrix for some time and then turns its direction clockwise to swirl around the inner separatrix.
As time elapses further, the trajectory falls either into the $L$- or the $R$-basin of attraction depending on the sign of the successive quasienergy change after the entry.
In Fig.~\ref{Fig6:Afterentering}, we present this finding from two selected initial states.
We observe in Fig.~\ref{Fig6:Afterentering},(a) that the trajectory that started out the initial point $(q_0,p_0)=(1.7,0)$ has fallen into the $L$-basin of attraction and continues relaxing to the $L$-point.
In contrast, in Fig.~\ref{Fig6:Afterentering}(b) we see that the initial state $(q_0,p_0)=(-1.0,-1.87)$ evolves into the $R$-basin of attraction.

Next, we analyze the dynamics upon `entering' the outer separatrix in great detail.
The Arnold problem addresses the question in the vanishing damping limit, $\gamma\rightarrow 0$.
In principle, the limiting trajectory would tend toward the separatrix on an extremely long time scale, which cannot be achieved numerically.
For the sake of argument, we take a portion of the outer homoclinic orbit as an approximate trajectory of motion just after the entry takes place.
To be more specific, let us denote $\tilde E_A$ as the quasienergy at point $A=(q_A,p_A)$ in Fig.~\ref{Fig3:separatrix2}, whose value equals $\tilde E_{sd}$.
After entering, the particle continually evolves along the aforementioned separatrix segment while gaining quasienergy by continuous counterclockwise motion around the outer loop and losing quasienergy with clockwise motion along the inner branch.
As the particle approaches the saddle point marked as $B$ in Fig.~\ref{Fig3:separatrix2}, its accumulated quasienergy becomes
\[\tilde E={\tilde E}_A+\delta {\tilde E}_{out}+\delta {\tilde E}_{in},\]
where $\delta {\tilde E}_{out}\equiv \delta {\tilde E}_{A\rightarrow B}$ represents the quasienergy gain along the elapsed segment of the outer loop, and $\delta {\tilde E}_{in}$ represents the quasienergy loss along the inner loop, which equals $-{\bar\gamma} S_R$ from Eq.~(\ref{innnerloss}).
If $\delta {\tilde E}_{out} > |\delta {\tilde E}_{in}|$, the arrival quasienergy of the particle at $B$ exceeds the saddle point value ${\tilde E}_{sd}$.
Accordingly, the particle must evolve into the $L$-basin of attraction to eventually reach the L-point.
This situation is depicted in Fig.~\ref{Fig6:Afterentering}(a).
On the other hand, if $\delta {\tilde E}_{out} <  |\delta {\tilde E}_{in}|$, the net change in quasienergy $\delta {\tilde E}_{in}+\delta {\tilde E}_{out}$ becomes negative so that the quasienergy of the particle as it approaches the saddle point becomes less than the actual saddle point energy ${\tilde E}_{sd}$.
Then, the particle trajectory must fall into the $R$-basin of attraction, as depicted in Fig.~\ref{Fig6:Afterentering}(b).
Thus, the states entering the separatrix are categorized into two classes based on the quasienergy position $A_c$ at which the net-quasienergy change balances, $\delta {\tilde E}_{out}+\delta {\tilde E}_{in}=0$; so, $\delta {\tilde E}_{A_c\rightarrow B}={\bar\gamma} S_R$.
The states entering the lower segment ($A_c\rightarrow B$) will be attracted to the $R$-basin with an accumulated loss of quasienergy $\delta {\tilde E}_{out}+\delta {\tilde E}_{in}<0$.
The states entering into the upper segment ($B\rightarrow A_c$) will gain more quasienergy than ${\bar\gamma}S_R$ and consequently accumulate a net quasienergy $\delta {\tilde E}_{out}+\delta {\tilde E}_{in}>0$ to fall into the $L$-basin of attraction.
Recall that the quasienergy change along the complete loop, $B\rightarrow A_c$ plus $A_c\rightarrow B$, gives the full area enclosed by the outer homoclinic orbit, ${\bar\gamma}(S_L + S_R)$ [Eq.~(\ref{outergain})].
Accordingly, because $\delta {\tilde E}_{A_c\rightarrow B}={\bar\gamma} S_R$, the quasienergy change $\delta {\tilde E}_{B\rightarrow A_c}$ must be ${\bar\gamma} S_R$.
The desired bifurcation probability is known to be proportional to the quasienergy changes \cite{Haberman1995}.
Thus, the probability of an initial state, which enters either loop segment, subsequently relaxing to either steady state, is proportional to the area of the corresponding basin of attraction, i.e., $S_L$ or $S_R$.
\[P_\alpha = \frac{{\bar\gamma}S_\alpha}{{\bar\gamma}(S_L+S_R)}\rightarrow \frac{S_\alpha}{S_L+S_R}, \quad \alpha=L,\ R,\]
which takes the same form as Eq.~(\ref{ArnoldFormula}).
Our analysis has revealed that the Arnold formula holds in the classical Josephson dynamics.

\section{Quantum dynamics in the Josephson junction}
\label{Quantum-dynamics}
Here, we consider Josephson junctions on a mesoscopic scale, where the Josephson phase is expected to behave as a quantum mechanical object.
We work in the parameter range, where the driving frequency is smaller than the zero-bias plasma frequency and the magnitude of the bias current is fixed.
Accordingly, neither the transition of the phase particle into excited levels nor the tunneling out of the effective potential well by sweeping current is of our concern.
We shall focus on the temporal development of the Arnold bifurcation for individual levels injected near the classical separatrix.
As far as we know, this type of manifestation of dynamical realization of the Arnold problem in a quantum system has not been previously considered in the literature.

\subsection{Single electron-pair spectrum}
\label{single-particle spectrum}
To consider the Arnold problem quantum mechanically, we first translate the dynamical variables in the classical Hamiltonian given in Eq.~(\ref{cHam}) into the Hermitian operators
\[
\varphi\rightarrow \hat \varphi \quad{\rm and}\quad N = \frac{1}{2e}C\bar\Phi_0{\dot \varphi}\rightarrow \hat N,
\]
where $N=Q/(2e)$ is the difference in the number of Cooper pairs, $Q$ being the net charge stored across the Josephson junction.
The operators $\hat\varphi$ and $\hat N$ are canonically conjugate with each other to satisfy the commutator $[\varphi,\hat{N}]=i$, which allows us to introduce the creation $\hat a^\dagger$ and annihilation $\hat a$ operators satisfying $[\hat a, \hat a^\dagger]=1$.
Consequently, we obtain the Duffing Hamiltonian as
\begin{widetext}
\begin{eqnarray}
\label{qHam2}
\hat H &=& \hbar\omega_0 ( \hat a^\dagger \hat a + \frac{1}{2}) -\varepsilon (\hat{a}+\hat{a^\dagger})^4- f_0 \cos(\omega t)(\hat{a}+\hat{a^\dagger}) - E_J,
\end{eqnarray}
\end{widetext}
where $\hbar\omega_0\equiv\sqrt{2 E_CE_J}$ with $E_C=(2e)^2/2C$ being the charging energy per electron pair.
In addition, the definitions of $\varepsilon$ and $f_0$ are related with the parameters $\beta$ and $f$ in the classical Hamiltonian as
\begin{eqnarray*}
&&\varepsilon\equiv\frac{1}{48}E_C=\frac{1}{3}\hbar\omega\sqrt{\frac{E_C}{2E_J}}\beta,\\
&&f_0\equiv I_0 \bar\Phi_0(\frac{E_C}{2 E_J})^{1/4}=\hbar\omega(\frac{2E_J}{E_C})^{1/4}f.
\end{eqnarray*}
The obtained Hamiltonian is time-dependent with periodicity of the driving current $\omega/2\pi$, i.e., ${\hat H}(t)={\hat H}(t+\omega/2\pi)$.

Here, we find it useful to define the unitary transformation,
\begin{equation}\label{unitary1}
\mid\Psi\rangle\rightarrow \mid\Psi_{RWA}\rangle=\hat{U}\mid\Psi\rangle,
\end{equation}
where $|\Psi\rangle$ and $|\Psi_{RWA}\rangle$ are the kets governed by the original $\hat H$ and the transformed $\hat{H}_{RWA}$, respectively,
and the unitary operator $\hat U$ is defined as
\begin{equation}
\label{unitary2}
\hat{U}=e^{i\omega \hat{a^\dagger}\hat{a} t},
\end{equation}
which is a quantum version of the classical parametrization represented by Eqs.~(\ref{trans1}) and (\ref{trans2}).
Subsequently, one can show that the two Hamiltonians are related to each other in accordance with
\begin{equation}
\label{trqHam}
\hat{H}_{RWA}=\hat{U}\hat{H}\hat{U}^{\dagger}-\hbar\omega \hat{a^\dagger}\hat{a},
\end{equation}
and the quantum dynamics is described by the time-dependent Schr\"odinger equation in the rotating-wave frame
\begin{equation}
\label{Schr-RWA}
i\hbar \frac{\partial }{\partial t}\mid\Psi_{RWA}\rangle = \hat{H}_{RWA} \mid\Psi_{RWA}\rangle.
\end{equation}
Next, as we performed the SVAA in Sec.~\ref{Classical-dynamics}, we take the average of the preceding $\hat{H}_{RWA}$ over the half-period $\pi/\omega$ of the driving current.
Consequently, we obtain the coarse-grained Hamiltonian for quantum analysis of the Arnold problem up to a constant as
\begin{equation}
\label{qHam3}
\hat{H}_{RWA}\rightarrow\hat{H}_{rwa}\equiv \hbar\bar\omega_0\hat{a^\dagger}\hat{a}-6 \varepsilon(\hat{a^\dagger}\hat{a})^2-\frac{f_0}{2}(\hat{a^\dagger} +\hat{a}),
\end{equation}
where we have set $\hbar\bar\omega_0\equiv \hbar(\omega_0-\omega) -6 \varepsilon$.
Note that all the oscillatory time-dependencies have been smoothed out in Eq.~(\ref{qHam3}) and the Hamiltonian $\hat{H}_{rwa}$ becomes approximately time-independent, or more precisely, \textit{quasi-autonomous}.

Thus, the system is in the quasi-stationary state within the slowly varying rotating wave approximation (RWA) and is described, in general, as
\begin{equation}
\label{RWAstationary}
\mid\Psi_{rwa}(t)\rangle = \sum_j a_j e^{-\frac{i}{\hbar}\tilde E_j t}\mid\phi_j\rangle,
\end{equation}
where $\tilde E_j$ and  $|\phi_j\rangle$ are solutions of the energy-eigenvalue equation,
\begin{equation}
\label{eigen1}
\hat H_{rwa}\mid\phi_j\rangle = \tilde E_j\mid\phi_j\rangle.
\end{equation}
We shall call the quasi-stationary energy $\tilde E_j$, defined on the coarse-grained time scale, the quasienergy.
Conventionally, the term `quasienergy' is attributed to the situation where a periodically driven Hamiltonian is considered strictly \cite{Shirley1965,Zeldovich1966,Ritus1966}.

We have solved the time-independent Schr\"odinger equation, Eq.~(\ref{eigen1}) in the number representation (Fock basis),
\begin{equation}\label{num-rep}
\mid\phi_j\rangle = \sum c_n^{(j)}\mid n\rangle,
\end{equation}
where $|n\rangle$ represents the eigenket of the number operator $\hat n=\hat{a^\dagger}\hat{a}$ and $c_n^{(j)}$ is the expansion coefficient.
In doing so, we have used the following numerical values adopted from \cite{Siddiqi2004},
\[
\omega=0.878 \omega_0,\quad \varepsilon=3.28\cdot 10^{-5}\hbar\omega_0,\quad{\rm and}\quad  f_0=0.89\hbar\omega_0,
\]
which are equivalent to the parameters used in Sec.~\ref{Classical-dynamics}.
The natural frequency for the $\rm {Al/Al_2O_3/Al}$ tunnel junction is estimated as $\omega_0\doteq 11.3~{\rm GHz}$, which gives a temperature scale $\hbar\omega_0/k_B\doteq 0.1K$.
Moreover, the charging energy per Cooper pair and the Josephson energy are estimated to be $E_C\doteq 1.2\times 10^{-5}~{\rm meV}$ and $E_J\doteq 2.4~{\rm meV}$.

\begin{figure}[h!]
\begin{center}
\includegraphics[width=0.4\textwidth]{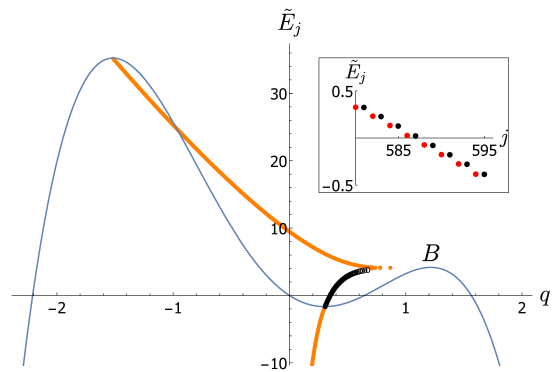}
\caption{[Color online] Quasienergy eigenvalues in the RWA versus the expectation values of the Josephson phase $\langle\varphi\rangle=q$ at the corresponding states; the contour of the classical quasienergy surface for fixed $p=0$ is inserted for reference (see Fig.~\ref{Fig2:Esurface}(b)). In addition, the notation B indicates the level corresponding to the saddle point. Note that $\tilde E_1$ corresponds to $\tilde E_{L}$, $\tilde E_{487}$ is closest to $\tilde E_{sd}$, and the paired $\tilde E_{618}$ and $\tilde E_{619}$ are closest to $\tilde E_{R}$. The inset shows a portion of the degenerated eigenvalues in the energy-index window $488\le j\le 619$, which is located inside the $R$-well, where the red and black dots are the localized and extended states, respectively.}
\label{Fig7:Espectrum}
\end{center}
\end{figure}
The result is given in Fig.~\ref{Fig7:Espectrum}, where we present the quasienergy spectrum as a function of the average phase, $\langle\varphi\rangle_j = \langle \phi_j|\hat \varphi|\phi_j\rangle$, of the Josephson junction at quantum level $j$, which corresponds to the slowly varying amplitude $q$ in classical dynamics.
In classical dynamics, the quantum average must be replaced with the time average. The time-average of the classical representation given in Eq.~(\ref{trans1}) for $\varphi$ over the symmetric half-period, $-\pi/2\omega\le t\le \pi/2\omega$, yields $\langle\varphi\rangle=q$, which is the average amplitude.
It is seen that, as we follow the energy levels from below, the corresponding stationary position $q$ of the particle moves from the origin toward the bottom of the $R$-well and continues in the positive direction but does not quite reach the saddle point.
Then, it changes its direction and continues climbing up to reach the $L$-point at the highest level (${\tilde E}_1$).
The states plotted with black color inside the well, $\tilde E_{R}\le {\tilde E}_j\le \tilde E_{sd}$, cover the RWA levels $488\le j\le 619$.
Furthermore, intriguingly, they are nearly doubly-degenerate, as can be seen the inset.
The closest levels to $\tilde E_{R}$ inside the $R$-well are $\tilde E_{618}$ and $\tilde E_{619}$ belonging to the inner lobe and outside the large separatrix in Fig.~\ref{Fig3:separatrix2}, respectively.
We have numerically checked that increasing the number of basis vectors simply produces more negative energies without affecting the upper branch in the spectrum.
Note that the energy spectrum of the quasi-Hamiltonian $\hat H_{rwa}$ is bounded from above, and not from below.

Having solved the problem in the RWA, we now consider the dynamics governed by the original Hamiltonian $\hat H$ by performing the inverse transformation of Eq.~(\ref{unitary1}), which is performed approximately as $|\Psi\rangle \approx \hat{U}^\dagger\mid\Psi_{rwa}\rangle$ in the SVAA.
The temporal development of the state is then given by
\begin{equation}
\label{time-evol}
\mid\Psi(t)\rangle = e^{-i\omega {\hat a}^\dagger \hat a t}e^{-\frac{i}{\hbar}\hat H_{rwa}t}\mid\Psi(0)\rangle,
\end{equation}
where $|\Psi(0)\rangle$ is an arbitrary initial state that may be expanded in the rotating-wave basis $|\phi_j\rangle$ as
\[
\Psi(0)=\sum a_j\mid \phi_j\rangle.
\]
For later analysis, we shall recast the above Eq.~(\ref{time-evol}) into
\begin{equation}\label{inv-unitary1}
\mid\Psi(t)\rangle = \sum_ja_j e^{-\frac{i}{\hbar}{\tilde E}_jt}\mid \psi_j(t)\rangle,
\end{equation}
where we have used the expansion given in Eq.~(\ref{num-rep}) to define
\begin{equation}
\label{Fqbasis}
\mid\psi_j(t)\rangle\equiv e^{-i\omega {\hat a}^\dagger \hat a t}\mid\phi_j\rangle  = \sum_n c_n^{(j)} e^{-in\omega t}\mid n\rangle,
\end{equation}
which satisfies the periodicity of $\hat H$, \[\mid\psi_j(t+2\pi/\omega)\rangle = \mid\psi_j(t)\rangle.\]
In particular, if the system is prepared initially in an RWA eigenstate, say $|\Psi(0)\rangle=|\phi_j\rangle$,
it evolves simply as
\begin{equation}
\label{Floquet}
\mid\Psi(t)\rangle \rightarrow e^{-\frac{i}{\hbar}{\tilde E}_jt}\mid \psi_j(t)\rangle \equiv \mid\Psi_{ E_j}(t)\rangle.
\end{equation}
Then, the energy $E_j$ of the system in the quasi-stationary state $|\Psi_{E_j}\rangle$ may be still defined as the expectation value, $E_j\equiv\langle\Psi_{E_j} |\hat H|\Psi_{E_j}\rangle$, which can be calculated using Eq.~(\ref{trqHam}) as
\begin{eqnarray*}
&&\langle\Psi_{{E}_j} \mid (\hat U^\dagger \hat H_{rwa} \hat U + \hbar\omega {\hat a}^\dagger \hat a)\mid \Psi_{{E}_j}\rangle \nonumber\\
&=& \langle\Psi_{rwa} \mid \hat H_{rwa}\mid \Psi_{rwa}\rangle_j + \hbar\omega \langle \Psi_{{E}_j}\mid {\hat a}^\dagger \hat a \mid \Psi_{{E}_j}\rangle \nonumber\\
&=& {\tilde E}_j + \hbar\omega\sum_n n|c_n^{(j)}|^2.
\end{eqnarray*}
Thus, we get the quasi-stationary energy $E_j$ associated with the state $|\Psi_{{E}_j}\rangle$ as
\begin{equation}
\label{realE}
E_j =  {\tilde E}_j + \hbar\omega\sum_n n|c_n^{(j)}|^2.
\end{equation}
Note that we use the notation $E_j$ to denote the energy expectation value of $\hat H$, which is distinct from ${\tilde E}_j$ for the RWA energies.
The obtained approximate energy eigenvalues of $\hat H$ are time-independent and the states of the system evolve in the quasi-stationary manner on a coarse-grained time scale longer than $\pi/\omega$.

\begin{figure}[h!]
\begin{center}
\includegraphics[width=0.4\textwidth]{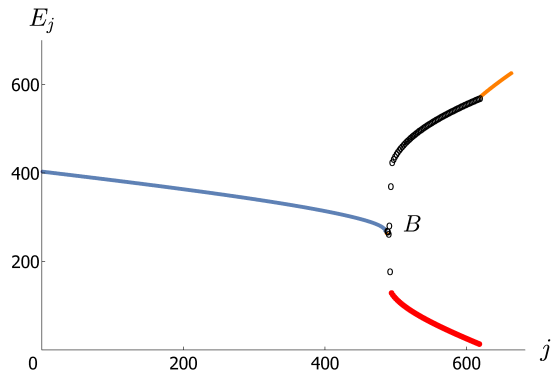}
\caption{[Color online] The quasienergy spectrum of $\hat H$ in units of $\hbar\omega_0$, where an intriguing feature is seen, namely, energies are not monotonically distributed with the quantum number $j$. The split black and red branches, $488\le j\le 619$, correspond to the degenerate levels inside the well in the classical energy contour in Fig.~\ref{Fig7:Espectrum}. The notation B indicates the level corresponding to the saddle point.}
\label{Fig8:Realspectrum}
\end{center}
\end{figure}
We present the energy spectrum of $\hat H$ given by Eq.~(\ref{realE}) in Fig.~\ref{Fig8:Realspectrum}, which is bounded from below in contrast to that of $\hat H_{rwa}$.
Because of the contribution of the second term on the RHS in Eq.~(\ref{realE}), the quasi-eigenvalue corresponding to each level $j$ has been changed in comparison with Fig.~\ref{Fig7:Espectrum}.
The splitting structure in the energy band $488\le j\le 619$ arises from the nearly degenerated levels, which occupy the RWA-energy well alternatively in Fig.~\ref{Fig7:Espectrum};
the second term on the RHS in Eq.~(\ref{realE}) produces distinct contributions to two nearly degenerate states.
Consequently, the quasi-degeneracy in the $\hat H_{rwa}$ spectrum is lifted to result in two distinctive branches in the original $\hat H$ spectrum.
The highest level $j=1$ in Fig.~\ref{Fig7:Espectrum} is not the highest level anymore in Fig.~\ref{Fig8:Realspectrum}.
Moreover, the lowest level $j=618$ on the red-colored branch in the energy spectrum gives the ground state of the physical Hamiltonian $\hat H$.
Its paired partner $j=619$ near the bottom of the $L$-well in the classical energy contour in Fig.~\ref{Fig7:Espectrum} appears on the upper (black-colored) branch with a larger energy.
In addition, the sparse levels near the notation B are the ones that approach close to the saddle point in Fig.~\ref{Fig7:Espectrum}.
The two pronounced levels at $j=488$ and $j=489$ near B correspond to the highest degenerate states in the RWA well in Fig.~\ref{Fig7:Espectrum}.
For reference, we have checked that the level spacing between adjacent levels on the red branch is in the order of $\hbar\omega_0$.
Moreover, we have drawn the energy levels only up to $j=650$ by reflecting the quadruple-order expansion of the potential operator.

\begin{figure}[h!]
\begin{center}
\includegraphics[width=0.4\textwidth]{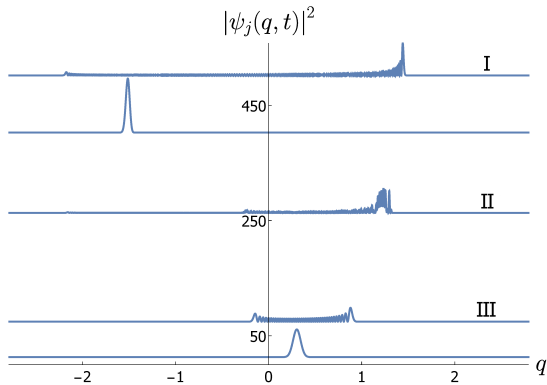}
\caption{Some of the quasi-stationary wavepackets at arbitrary values of time when their average locations are maximally departed from the center. All of these oscillate back and forth about the center with periodicity of $2\pi/\omega$, where the vertical axis indicates the energy scale defined in Fig.~\ref{Fig8:Realspectrum}.}
\label{Fig9:Wavepacket}
\end{center}
\end{figure}
Next, we examine the time-evolution of the quasi-stationary state $|\Psi_{E_j}(t)\rangle$ (see Eq.~(\ref{Floquet})).
The corresponding wavefunction is given explicitly in $q$-representation as
\begin{equation}
\label{realWF}
\Psi_{ E_j}(q,t) = e^{-\frac{i}{\hbar}{\tilde E}_jt} \psi_j(q,t),
\end{equation}
where
\[
\psi_j(q,t) = \sum_n c_n^{(j)} e^{-i\omega n t}H_n(q),
\]
where $H_n(q)\equiv \langle q| n\rangle$ are the eigenfunctions of the harmonic oscillator.

In Fig.~\ref{Fig9:Wavepacket}, we have illustrated the squared amplitude of the wavefunctions $\left|\Psi_{E_j}(q,t)\right|=\left|\psi_j(q,t)\right|$ for a few levels.
The pronounced characteristic of the quasi-stationary states is that their wave functions oscillate resonantly with the external frequency $\psi_j(q,t)=\psi_j(q,t+2\pi/\omega)$ while the shape of the wave packets are modulated over the period.
The two particularly well-localized wavepackets in Fig.~\ref{Fig9:Wavepacket} are from the levels $j=1$ and $j=618$ in Fig.~\ref{Fig8:Realspectrum}, which correspond to the top of $L$-hill and the bottom of $R$-well in the classical RWA energy contour, respectively, as shown in Fig.~\ref{Fig7:Espectrum}.
The wavepacket marked as II is the one corresponding to $j=487$, marked as B, which is mostly close to the saddle point in the classical energy contour.
The wavepackets marked as I and III are from the two split states, $j=537$ and $536$ in Fig.~\ref{Fig8:Realspectrum}, which, in turn, are the degenerate levels in Fig.~\ref{Fig7:Espectrum}.
In fact, all the paired levels, one depicted as black and its partner as red branches in Fig.~\ref{Fig8:Realspectrum}, which are split from the corresponding nearly-degenerate levels inside the RWA energy well, manifest the same feature, i.e., one level is delocalized and the other is localized.
In the classical picture, the localized levels with nearly same RWA energies ${\tilde E}_j$ in the window $\tilde E_{R}\le {\tilde E}_j\le \tilde E_{sd}$ belong to the inner lobe, and the delocalized levels belong to the region outside the large lobe in the separatrix.
Exceptionally, we found that the paired split levels of (488,489), (490,491), and (492, 493), which are inside the well and very close to ${\tilde E}_{sd}$, exhibit all extended states.

\subsection{Dissipative Josephson dynamics in a boson bath}
\label{DMformulation}
So far, we have considered the single-particle spectrum of the Josephson particle and quasi-stationary dynamics in a pure state without dissipation.
We now suppose that the Josephson atom described by $\hat H$, given in Eq.~(\ref{qHam2}), is placed in a heat reservoir at an absolute temperature $T$, whose value is low enough to satisfy the quantum criterion $\hbar\omega_0\gg k_BT$.
We express the total Hamiltonian for the composite system of the Josephson junction and the reservoir as
\begin{equation}
\label{eq:rwa41}
\hat{H}_{tot}=\hat{H}+\hat{H}_R+\hat{V},
\end{equation}
where $\hat{H}_R$ is the Hamiltonian of the reservoir composed of a number of bosonic modes;
\begin{equation}
\label{ResHam}
\hat{H}_R = \sum_{i}  \hbar \Omega_i \hat{b}_i^+ \hat{b}_i,
\end{equation}
and $\hat{V}$ is the interaction between the system and the reservoir, which can be simply modeled as
\begin{equation}
\label{eq:rwa43}
\hat{V} = (\hat{a}^++\hat{a}) \sum_i \kappa_i(\hat{b}^+_i + \hat{b}_i),
\end{equation}
where $\kappa_i$ is the coupling constant between the Josephson particle with $i$th bath mode.
Therefore, our model introduces dissipation via the system-reservoir interaction in the independent particle picture.
Because of the interaction, the Josephson particle is in a mixed state.

Here, we argue that 1) the coupling between the system and the reservoir is very weak; and 2) the reservoir remains in equilibrium with tremendously large degree of freedom.
Then, by performing the standard Born--Markov approximation \cite{Blum1981,Carmichael2002}, we obtain the Pauli master equation for the dissipative dynamics of the Josephson particle as
\begin{eqnarray}
\label{Pauli}
\dot P_j = \sum_{l \neq j} ( W_{jl} P_l - W_{lj}P_j),
\end{eqnarray}
where $P_j$ is the occupation probability of an arbitrary RWA level $|\phi_j\rangle$.
The details of the nontrivial steps in the derivation are given in the Appendix.
In Eq.~(\ref{Pauli}), $W_{jl}$ is the transition rate from state $l$ to $j$, which we have identified as
\begin{equation} \label{trate}
W_{jl}= \frac{2\pi\kappa^2}{\hbar^2} \Big( |a_{jl}|^2 {\bar W}_{jl}
+ |a_{lj}|^2 {\bar W}_{lj}\Big),
\end{equation}
where, for notational convenience, we have introduced
\begin{widetext}
\begin{eqnarray*}
{\bar W}_{jl}&\equiv& g(-\omega_{jl}+\omega)(\bar{n}(-\omega_{jl}+\omega)+1) + g(\omega_{jl}-\omega)\bar{n}(\omega_{jl}-\omega), \\
{\bar W}_{lj} &\equiv& g(\omega_{lj}-\omega)(\bar{n}(\omega_{lj}-\omega)+1) + g(-\omega_{lj}+\omega)\bar{n}(-\omega_{lj}+\omega).
\end{eqnarray*}
\end{widetext}
In the preceding definitions, the arguments in the expressions for the density of bosonic modes $g$ and Planck distribution $\bar n$ must be positive.

Next, we solve Eq.~(\ref{Pauli}) numerically for the same physical parameters used in Sec.~\ref{single-particle spectrum} in a wide range of initial conditions for the given coupling constant $\kappa$ and temperature $T=(k_B\beta)^{-1}$.
The temperature is embedded in the formulation via the Planck distribution of the reservoir modes.
For numerical purposes, we make the time and coupling constant dimensionless according to
\[t^* = t/t_p \quad{\rm and}\quad \kappa^* = \frac{\kappa}{\hbar\omega_0},\]
where $t_p$ is defined as $t_p \equiv\frac{\pi c_s^3}{V \omega_0^2\omega^2}$.
Then, Eq.~(\ref{Pauli}) is reduced to a dimensionless form,
\[
\frac{dP_j}{dt^*} = {\kappa^*}^{2}\sum_{l \neq j} ( W_{jl}^* P_l - W_{lj}^*P_j),
\]
where the dimensionless transition rate $W_{jl}^*$ follows from Eq.~(\ref{trate}), but is not presented.
Note that the quadratic dependence on the coupling constant stems from our second-order Born approximation in handling the interaction between the system and the reservoir (see Eq.~(\ref{IntSysEq3})).
Throughout the calculation, we have checked the probability conservation as a consistency condition $\sum P_j(t)=1$.

\begin{figure}[h!]
\begin{center}
\includegraphics[width=0.42\textwidth]{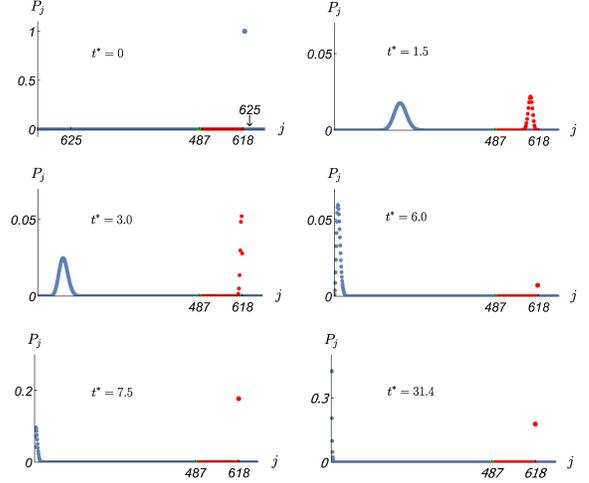}
\caption{Time-evolution of the quantum occupancy $P_j$  when $\kappa^*=1$ at zero temperature, $T=0$; we have indicated the initial state with occupancy one at $j=625$ by a vertical arrow and the unit of time is $t_p$ given in the main text. The quantum level at $j=487$ is closest to the classical saddle point energy ${\tilde E}_{sd}$. At the longest time shown at $t^*=31.4$, the partially-occupied levels near $j=1$ and the isolated $j=618$ on the right form the steady-state distribution.}
\label{Fig10:QP625}
\end{center}
\end{figure}
In Fig.~\ref{Fig10:QP625}, we present the temporal-development of the quantum occupancy $P_j$ at several time steps, where the system was initially prepared at energy level $j=625$ (see Fig.~\ref{Fig8:Realspectrum}).
This initial level lies below the bottom energy, ${\tilde E}_{R}$, of the classical RWA well in Fig.~\ref{Fig7:Espectrum} and corresponds to a phase point outside the large separatrix in Fig.~\ref{Fig3:separatrix2}.
We observe that the initially empty levels below the injection level get gradually excited as time elapses via the interaction with reservoir.
Note that the states on the black-colored branch in the energy spectrum (see Fig.~\ref{Fig8:Realspectrum}) have been excited but have already decayed to empty on the time scales we have shown.
In the long time limit, the system tends to become an incoherent mixture of only a small number of stationary states, a few levels near $j=1$, and the well isolated level, $j=618$.

\begin{figure}[h!]
\begin{center}
\includegraphics[width=0.45\textwidth]{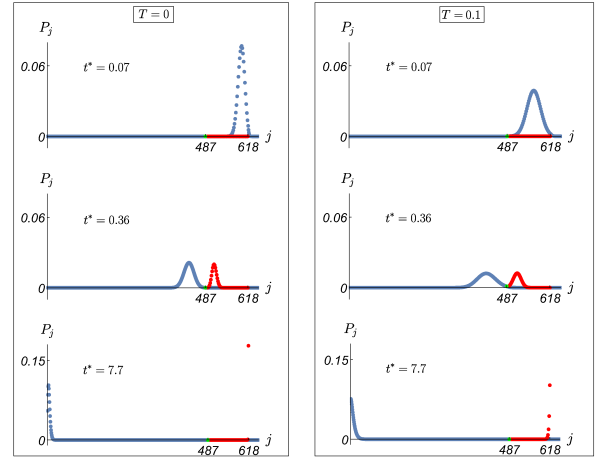}
\caption{Comparison of time-evolutions of the occupation probability distributions at two different temperatures, $T=0$ and $0.1$ in Kelvin, for a fixed $\kappa^*=1$; in both the cases, the system is excited at the energy level $j=625$ and subsequent dynamics follows at three time steps.}
\label{Fig11:QPtemp}
\end{center}
\end{figure}
We have further examined the occupancy dynamics from other initial conditions.
1) We assume initial injections at $j=500,\ 501$, which are nearly degenerate levels in the RWA spectrum, and thus correspond to two split levels in the original energy spectrum (see Fig.~\ref{Fig8:Realspectrum}), where $j=500$ belongs to the lower (red) branch and $501$ belongs to the upper (black) branch.
For injection at $j=500$, we observe that only `localized' levels in the energy window $488\le j\le 619$ get excited after the initial excitation, and the system eventually relaxes to a pure state at $j=618$ as $t\rightarrow \infty$.
On the other hand, for injection at $j=501$, we observe similar features as in Fig.~\ref{Fig10:QP625}.
2) We follow the occupation dynamics starting from the initial injection at a level, e.g., $j=400$, with energy ${E}_j < {E}_{487}$, which classically belongs to the shaded area in Fig.~\ref{Fig3:separatrix2}.
This initial state only relaxes to the left of $j=487$, and not the levels to the right, which appears to be in contrast to the classical case. This indicates a bistable feature depending on the damping strength.
This is because in the quantum case, the master equation was derived within the second-order Born approximation; thus, we consider the system only within the small damping limit.
As $t\rightarrow \infty$, the system tends to become an incoherent mixture of only the steady-state levels near $j=1$.
An exceptional detail is that the particular levels $j=484,\ 485,\ {\rm and}\ 486$ near the level $j=487$ excite the levels both to the left and to right of the level $j=487$ although ${E}_j < {E}_{487}$.

Next, we consider the temperature effect on the time-evolution of the occupancy distribution for a fixed initial condition.
The transition rate between two levels depends on temperature through the Planck distribution $\bar n$ of the reservoir modes (see Eq.~(\ref{trate})).
This must affect the dynamics of quantum occupancy distribution.
The numerical outcome is presented in Fig.~\ref{Fig11:QPtemp}, where the initial condition for both cases is chosen at $j=625$.
The general tendency of the occupation relaxation is the same at both temperatures.
In addition, the instantaneous dispersion of the occupancy distribution is wider at finite temperature, meaning that more levels are involved in the relaxation at a given time.

\begin{figure}[h!]
\begin{center}
\includegraphics[width=0.45\textwidth]{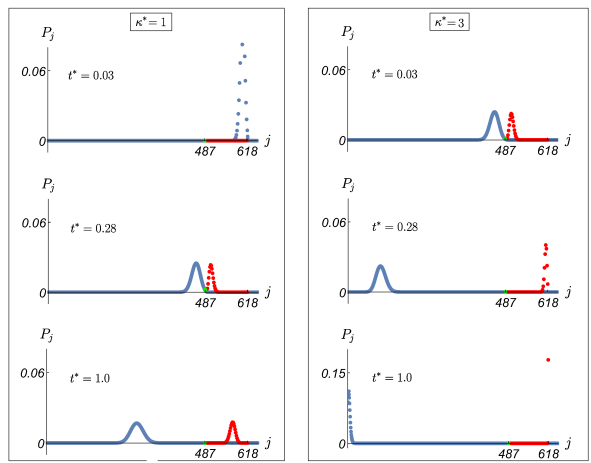}
\caption{Comparison of time-evolutions of the occupation probability distributions for two different damping strengths, $\kappa^*=1$ and $3$, at a fixed temperature $T=0$.}
\label{Fig12:QPkappa}
\end{center}
\end{figure}
In addition, in Fig.~\ref{Fig12:QPkappa}, we depict the level dynamics for two coupling constants between the system and the reservoir. Apparently, the relaxation to the steady-state distribution is faster when the coupling constant is stronger, which plays the role of intrinsic resistance in our Josephson junction model.

The preceding numerical results manifest that a chosen initial state $P_j(0)$ with its energy eigenvalue ${E}_j\ge E_{619}$ (see the energy spectrum, Fig.~\ref{Fig8:Realspectrum}) evolves into a mixed state of many levels below ${E}_j$.
As time elapses, there appears a time scale $t^*_c$ for a given damping magnitude $\kappa^*$ beyond which only well-localized levels, namely the red branch in the spectrum in the index window $488\le j\le 619$ and the blue branch in the spectrum at the levels below $j=487$ are involved in the population dynamics.
Moreover, the occupation probabilities of other levels, namely, the alternating degenerate levels, which are delocalized, with the localized levels in the same index window and the levels $j\ge 619$, remain empty.
For convenience, we categorize the localized states in the quantum-index window as `R' levels, forming a `quantum $R$-basin', and the states below ${E}_{487}$ as `L' levels, forming a `quantum $L$-basin'.
Here, $L$ and $R$ indicate the location of a chosen level with respect to the reference level ${E}_{487}$ (see Fig.~\ref{Fig10:QP625}).
Again, here, the levels $j=484,\ 485,\ {\rm and}\ 486$ represent an exceptional case wherein they do not belong to the $L$-basin although their energies are smaller than $j=487$.
The $L$-levels and $R$-levels correspond to the blue and red branches, respectively, in the energy spectrum in Fig.~\ref{Fig8:Realspectrum}.
Our numerical experiments reveal that after $t^*\ge t^*_c$, all $L$-levels except those in the exceptional cases relax down to the levels near $j=1$, and all $R$-levels relax onto the level $j=618$.
For an initial excitation at levels above $j=618$, the system reaches a steady state, ${\dot P}_j^{st}=0$, in the long-time limit for a given temperature and damping constant.
The steady-state distribution must hold the condition
\[\sum_{l \neq j} ( W_{jl} P_l^{st} - W_{lj}P_j^{st}) = 0, \]
where we have used the notation $P_j^{st}=P_j(\infty)$. The numerically obtained steady-state distribution $P_j^{st}$, for instance, the one shown at $t^*=31.4$ in Fig.~\ref{Fig10:QP625}, meets the above condition.
In addition, we have confirmed that the steady-state distribution $P_j^{st}$ satisfies the detailed-balance condition
\begin{equation}\label{balance}
W_{jl} P_l^{st} - W_{lj}P_j^{st} = 0.
\end{equation}

Next, we consider the q-representation of the density operator $\hat\rho$ given by
\[\rho(q,t)=\langle q|\hat\rho(t)|q\rangle,\]
which we interpret as the probability density of the Josephson particle being found in the range $(q,q+dq)$ at a given time $t$.
By utilizing the effective form for $\hat\rho$ in the reduced Hilbert space of the system alone,
\[
\hat\rho=\sum_{ij} \rho_{ij} \left| \Psi_{ E_i}(t)\right\rangle \left\langle \Psi_{E_j}(t)\right|,
\]
we calculate the probability density once again in the RWA basis.
Consequently, within the diagonal approximation, we obtain the ensemble density as an incoherent superposition of the probability occupancies $P_j$,
\begin{equation}
\label{q-Density}
\rho(q,t) = \sum_j P_j(t)|\psi_j(q,t)|^2,
\end{equation}
where the wavefunctions $\psi_j(q,t)$ appearing in the weighting coefficients have been specified in Eq.~(\ref{realWF}).
The $q$-representation of the RWA density operator is given by
\[\tilde\rho(q,t) = \sum_j P_j(t)|\phi_j(q)|^2,\]
which differs from Eq.~(\ref{q-Density}) by only the oscillatory factor $e^{-i\omega nt}$ in the quasi-eigenfunctions.
We have numerically confirmed that $\tilde\rho(q,t)\rightarrow\rho(q,t)$ at $t^*\simeq 31.4$, and its shape remains fixed beyond that time scale without further oscillation.

\begin{figure}[h!]
\begin{center}
\includegraphics[width=0.4\textwidth]{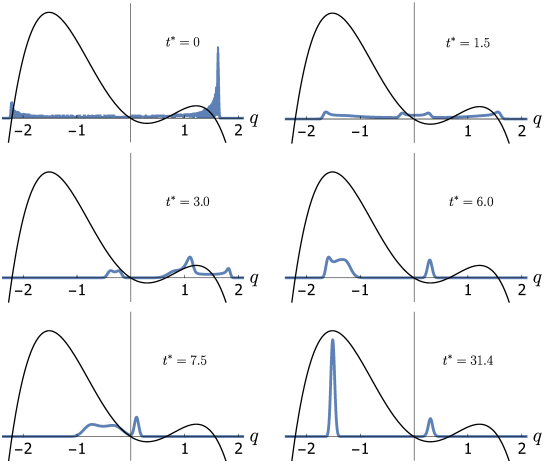}
\caption{Temporal evolution of the probability density $\rho(q,t)$ for initial excitation at energy level $j=625$ with $T=0$ and $\kappa^*=1$; for numerical purposes, we have set the values of time $t^*$ in units of $t_p\equiv\omega^{-1}$. The contour of the classical energy surface has been inserted for reference (see Fig.~\ref{Fig2:Esurface}(b)).}
\label{Fig13:Qdensity}
\end{center}
\end{figure}
In Fig.~\ref{Fig13:Qdensity}, we illustrate the development of the probability density with time.
The initial excitation at a particular energy level at $j=625$ results in a broadened probability density at $t^*=0$, reflecting the extended state $\langle q|\phi_{625}\rangle$.
As time elapses, other energy levels get excited, which results in an intriguing feature in the ensemble density.
In the steady-state limit, e.g., at the longest time $t^*=31.4$ shown in the figure, the probability density manifests a bimodal shape peaked at two positions to which the $L$-top and $R$-well respectively correspond in the classical energy contour.
In that time scale and beyond, the probability density evolves periodically according to
\[\rho(q,t) \rightarrow \sum_j P_j(\infty)|\psi_j(q,t)|^2,\]
where the summation is performed only over the steady-state levels (see Fig.~\ref{Fig10:QP625}).
We have numerically confirmed the quasi-stationarity of the density function at time $t^*\ge 31.4$, $\rho(q,t)=\rho(q,t+2\pi/\omega)$, reflecting the periodic behavior of the quasi-stationary wavefunctions $\psi_j(q,t)$ shown in Fig.~ \ref{Fig9:Wavepacket}.
We remark that the frequency-locked, adiabatic synchronization with the driving oscillation is reminiscent of \textit{autoresonance} in nonlinear systems \cite{Frieland2009}.
However, the underlying mechanism is different; in our case, no frequency chirping is used, and the amplitude does not increase \cite{Naaman2008}.

\begin{figure}[h!]
\begin{center}
\includegraphics[width=0.5\textwidth]{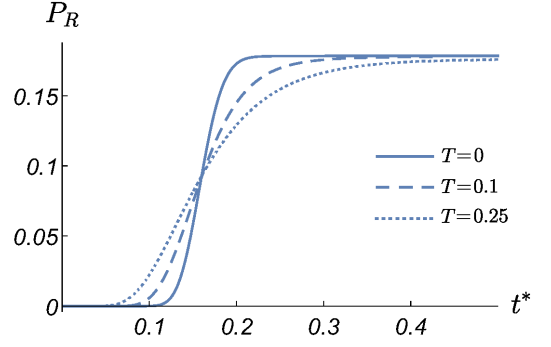}
\caption{The probability of finding the Josephson particle in the quantum $R$-basin for different temperatures $T$ in Kelvin with fixed $\kappa^*=1$; the initial condition was chosen at $j=625$ below the bottom of the $R$-well in the RWA. The threshold time is shortest at $T=0$, $t^*_c\doteq 0.25$, and it becomes longer as the temperature increases.}
\label{Fig14:QPRtemp}
\end{center}
\end{figure}
Now, to address the Arnold problem, we quantify the probability of eventually finding the system in the quantum $R$-basin, which is initially prepared at a level above $j=618$ in the energy spectrum, corresponding to an RWA level below the bottom of the classical energy well, as
\begin{eqnarray}
\label{ProR}
P_R(t^*)=\sum_{j=R} P_j(t^*).
\end{eqnarray}
Similarly, the conjugate probability of finding the particle in the quantum $L$-basin may be quantified as
\[
P_L(t^*)=\sum_{j=L} P_j(t^*).
\]
In Fig.~\ref{Fig14:QPRtemp}, we draw the numerical outcome of $P_R$ as a function of time for several different temperatures, where, at $t^*=0$, we have excited the system at level $j=625$.
We observe that the probability $P_R$ tends to be a constant value of $0.18$ at all considered temperatures.
Evidently, the temperature effect is to slow down the relaxation.
A similar tendency was observed for $P_L(t^*)$ but has not been illustrated.
For numerical consistency, we have checked that the sum of $P_L$ and $P_R$ tends to approach unity, i.e., $P_L(t^*)+P_R(t^*)\rightarrow 1$, as the time exceeds the threshold $t^*\ge t^*_c$.

\begin{figure}[h!]
\begin{center}
\includegraphics[width=0.5\textwidth]{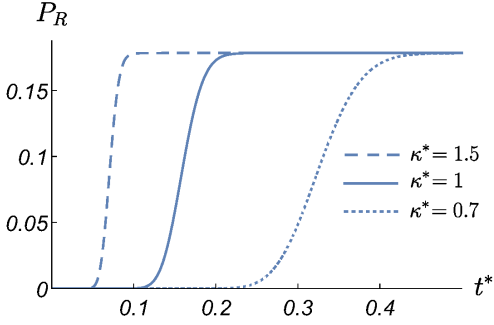}
\caption{The probability of finding the Josephson particle in the quantum $R$-basin at a fixed temperature $T=0$ for several damping strengths; the threshold values of time are estimated to be $t^*_c \doteq 0.11,\ 0.25, {\rm and}\ 0.50$ in decreasing order of the coupling constant $\kappa^*$.}
\label{Fig15:QPRcoupling}
\end{center}
\end{figure}

In addition, in Fig.~\ref{Fig15:QPRcoupling}, we show the effect of damping on $P_R$ for a fixed temperature.
One can see that the threshold or relaxation time gets smaller as the coupling constant $\kappa^*$ increases.
Although our working model, i.e., the master equation, is limited within the second-order Born approximation, the results agree with the general expectations.

Finally, we formulate the Arnold bifurcation probability in the quantum dissipative dynamics of the Josephson particle.
Our goal here is to derive a quantum analog of Eq.~(\ref{ArnoldFormula}) by finding the probability that an initial excited state with energy above the saddle point level ${E}_{487}$ is captured into either the quantum L- or the R- basin.
To this end, we choose an initial condition $P_j(0)=\delta_{jm}$ with $E_m>E_{618}$ for Eq.~(\ref{Pauli}) and let it follow the dynamics.
After time-evolution on the threshold-time scale $t_c$, the population dynamics would be well separated between the $L$-levels and $R$-levels, as discussed earlier.
Then, we formally integrate the master equation over $t_c$ to obtain
\[
P_j (t_c) = \delta_{jm} + \sum_{l\neq j} \left( W_{jl} \int_0^{t_c} P_l(t) dt - W_{lj} \int_0^{t_c} P_j(t) dt\right),
\]
where the summation $l$ on the RHS covers all quantum levels.
We then take a summation of $P_j(t_c)$ over only the $R$-levels to obtain
\begin{eqnarray*}
&&\sum_{j \in R} \Big(P_j (t_c)-\delta_{jm}\Big)\\
&=&\sum_{j \in R}  \sum_{l\neq j} \left( W_{jl} \int_0^{t_c} P_l(t) dt - W_{lj} \int_0^{t_c}P_j(t) dt\right),
\end{eqnarray*}
where the second term on the left hand side vanishes because $m\notin R$.
Next, by taking advantage of the identity
\[\sum_{j \in R}  \sum_{l \in R} \left( W_{jl} \int_0^{t_c} P_l(t) dt - W_{lj} \int_0^{t_c} P_j(t) dt\right)=0,\]
we reduce the preceding equation to
\begin{widetext}
\begin{equation}
\label{Pauli4}
P_R(t_c) = \sum_{j \in R}  \sum_{l\notin R}  W_{jl} \int_0^{t_c} P_l(t) dt - \sum_{j \in R}  \sum_{l\notin R} W_{lj} \int_0^{t_c} P_j(t) dt,
\end{equation}
\end{widetext}
where $P_R(t_c)$ has been defined in Eq.~(\ref{ProR}).
Because all levels in the $R$-set relax to the level $j=618$, $P_R(t_c)$ gives a measure of the probability of the Josephson particle relaxing to the bottom of the quantum $R$-basin.
Similarly, by summing up all levels belonging to the $L$-set, we can determine the probability for the particle to relax into the $L$-basin.
The result is
\begin{widetext}
\begin{equation}
P_L(t_c)
= \sum_{j \in L}\sum_{l\notin L}  W_{jl} \int_0^{t_c} P_l(t) dt - \sum_{j \in L}  \sum_{l\notin L} W_{lj} \int_0^{t_c} P_j(t) dt.
\end{equation}
\end{widetext}
We have performed a detailed numerical analysis and determined that not all the energy levels in the summation provide an appreciable contribution to either $P_L$ or $P_R$.
In fact, it is observed that the transition rate between two levels near the classical saddle-point energy alone results in an appreciable value (see Fig.~\ref{Fig8:Realspectrum}).
Further, we have numerically confirmed that the time-integrals of the probability occupancy of such levels, between which the finite transition rate occurs, are nearly constant as
\[ \int_0^{t_c} P_l(t) dt \doteq 2.5\times 10^{-3},\]
with the relative discrepancy among different levels less than $3\%$.
Consequently, the probability of the Josephson particle, prepared initially at a level above $j=618$ of relaxing in time into either the quantum L- or R- basin may be quantified as
\begin{equation}\label{QArnold}
P_\alpha = \frac{\tilde S_\alpha}{{\tilde S}_L+{\tilde S}_R},
\end{equation}
where $\alpha=L,\ R$.
In Eq.~(\ref{QArnold}), ${\tilde S}_\alpha$ is a metaphorical expression corresponding to the phase-space area $S_\alpha$ in the classical formula, Eq.~(\ref{ArnoldFormula}), which is specified as
\begin{equation}
\label{Qbasin}
\tilde S_\alpha \approx {\sum}^\prime_{j \in \{\alpha\}}{\sum}^\prime_{l\notin\{\alpha\}} ( W_{jl} - W_{lj}),
\end{equation}
where the prime indicates that the summation encompasses only the quantum levels near the saddle-point energy.
By substituting Eq.~(\ref{trate}) into Eq.~(\ref{Qbasin}), we obtain an explicit representation for $\tilde S_\alpha$,
\begin{widetext}
\begin{equation}
\label{quantum-area}
\tilde S_\alpha = \frac{2 \pi \kappa^2}{\hbar^2} {\sum}^\prime_{j \in \{\alpha\}}{\sum}^\prime_{l\notin\{\alpha\}} \Big\{ |\hat{a}_{jl}|^2 \Big(g(-\omega_{jl}+\omega)-g(\omega_{jl}-\omega)\Big)
+ |\hat{a}_{lj}|^2 \Big(g(\omega_{lj}-\omega)-g(-\omega_{lj}+\omega)\Big) \Big\},
\end{equation}
\end{widetext}
where, as emphasized earlier, the arguments in the expression for the density of states $g$ must be positive.
Note that the phase-space area $\tilde S_\alpha$ is found to be independent of temperature.

We have applied Eq.~(\ref{QArnold}) to numerically estimate that $P_R\doteq 0.18$, which is consistent with the results presented in Fig.~\ref{Fig14:QPRtemp} and \ref{Fig15:QPRcoupling}.
The mathematical formula given in Eq.~(\ref{QArnold}) is the quantum analog of the Arnold probability given in Eq.~(\ref{ArnoldFormula}).
Like in the classical case, the Arnold probability is independent of the damping constant $\kappa$.

\section{Summary and Conclusion}
\label{Conclusion}

We have investigated the Arnold problem which addresses the bistable stochasticity in the dissipative Josephson dynamics.
We have viewed the Josephson junction as an artificial particle, which we named the `Josephson particle'.
We have addressed the Arnold problem first in the classical regime, where the relative phase was treated as a macroscopic continuum variable, and then continued formulating it in the quantum regime, where the phase was quantized.
The primary results we obtained are summarized below.

In the classical regime, we have formulated the problem regarding an effective phase dynamics in terms of the parameterized amplitudes.
We have smoothed out the fast oscillatory time-dependence to make the dynamics approximately autonomous, i.e., the SVAA.
The attendant Hamiltonian was not separated into the conventional kinetic and potential energy terms, and accordingly resulted in an intriguing energy landscape manifesting two stable fixed points and a saddle point.
Subsequently, the state-space trajectories have been examined to study the build-up of the bifurcation dynamics in detail.
We have found that as time develops, the trajectories evolve into either the $L$-basin or $R$-basin of attraction, depending on where one chooses the initial states in the phase-space area defined by the separatrix.
When the damping is significant, any initial condition renders the ensuing trajectory to relax into the fixed point in the $R$-basin.
In the opposite case of small damping, we have explored the bifurcation dynamics for the chosen initial states placed outside the separatrix with energy near the saddle point.
Consequently, we have derived the Arnold formula for describing the probability of the Josephson particle being captured into either equilibrium basin, which is proportional to the corresponding enclosed phase-space area of the homoclinic orbit.

In the quantum regime, we have cast the classical Hamiltonian into the operator representation so that the problem entails the quantum dynamics of a Josephson particle.
We have first solved the single-particle problem to obtain the energy eigenvalues and quasi-stationary eigenstates within the quantum version of the RWA.
The obtained energy spectrum reveals an unusual feature due to the unconventional structure of the Hamiltonian.
Furthermore, the corresponding wave functions exhibit coherent oscillations with the periodicity of the driving current.
Then, we have placed the system in interaction with a heat reservoir, which provides a dissipation mechanism.
We have set up a quantum Liouville equation for the total density operator and performed the standard Born--Markov approximation to derive a Pauli master equation for the reduced density operator of the system alone.
Subsequently, we have numerically solved the master equation to obtain the time development of the probability occupancy of the quantum levels.
Like in the classical case, if we inject the system into a level near the saddle-point value, the ensuing dynamics shows a bifurcation feature wherein, after a threshold time, only the levels in the quantum $L$-basin and $R$-basin possess finite occupancy while preserving the respective net probability.
In the long-time limit, the system reaches the steady state specified by an incoherent combination of the occupied levels.
Two essential features of the steady-state distribution are 1) the detailed balance condition is satisfied, which we have numerically confirmed; 2) the autoresonance characteristic of frequency-locking with the driving force occurs in the long-time oscillation of the ensemble density.
In addition, the effect of temperature was seen to decrease relaxation, whereas the effect of the coupling strength was to speed up the relaxation.
Finally, we have proved that the Arnold formula holds in quantum dynamics as well, where the quantum analog of the phase-space area is given by the net gain of the in-and-out transition rates of the levels near the saddle-point, belonging to the respective quantum $L$-basin and $R$-basin.

To conclude, we have answered the Arnold question in the dissipative Josephson dynamics both classically and quantum mechanically within a consistent formulation, and have derived and numerically analyzed the bifurcation probabilities.
We hope that the new insight that we have provided regarding the manipulation of the quasi-stationary states will further enhance our understanding of the required long relaxation time, coherence time, nonlinearity, and tunability of the transition in superconducting circuits.

\acknowledgments
This work was supported by the NRFK (NRF-2015K2A1B8068583) and by the RFBR (Project No. 16-57-51045, No. 18-07-01206, and No. 16-07-01012 NIF) under the framework of an international cooperation program.
The Russian team also acknowledges the support in part from the Ministry of Education of the Russian Federation (Contract No. 3.3026.2017).

\section*{Appendix}
We describe here the derivation of the master equation, Eq.~(\ref{Pauli}), while emphasizing the nontrivial steps that are relevant to our purpose.

We first set up the quantum Liouville equation for the total system consisting of the system and the reservoir as
\begin{equation}
\label{qLiouville}
i\hbar \frac{\partial \hat\rho_{tot} (t)}{\partial t} = [\hat{H}_{tot}, \hat\rho_{tot} (t)],
\end{equation}
where $\hat\rho_{tot}$ denotes the density operator of the total system.
The reduced density operator for the system is defined by $\hat \rho(t) = {\rm Tr}_R\hat\rho_{tot}(t)$, where ${\rm Tr}_R$ denotes the trace over the reservoir states.
To proceed formulation, we find it convenient to introduce the transformation $\hat\rho_{tot}(t)\rightarrow \tilde{\rho}_{tot}(t)$ as
\begin{equation}
\label{int-pic}
\tilde{\rho}_{tot} (t)= e^{\frac{i}{\hbar}\hat{H}_R t} \hat{S}^\dagger(t)\hat\rho_{tot} (t) \hat{S}(t) e^{-\frac{i}{\hbar}\hat{H}_R t},
\end{equation}
where $\hat S$ is the time-evolution operator appearing in Eq.~(\ref{time-evol}),
\begin{equation*} \label{time-devel)}
\hat{S}(t) \equiv e^{-i \omega  \hat{a^\dagger}\hat{a} t} e^{-(i/\hbar) \hat{H}_{rwa} t}.
\end{equation*}
Then, by substituting Eq.~(\ref{int-pic}) into Eq.~(\ref{qLiouville}) followed by some manipulation, we convert Eq.~(\ref{qLiouville}) into the desired interaction picture as
\begin{equation}
\label{IntqLiouville}
i\hbar \frac{\partial {\tilde\rho}_{tot} (t)}{\partial t}=[\hat{V}_{I}(t), {\tilde\rho}_{tot} (t)],
\end{equation}
where $\hat{V}_{I}$ is the coupling term transformed as
\begin{equation} \label{IntV}
\hat{V}_{I}(t)\equiv e^{\frac{i}{\hbar}\hat{H}_R t} \hat{S}^\dagger(t) \hat{V} \hat{S}(t) e^{-\frac{i}{\hbar}\hat{H}_R t}
=\hat{x}_I \sum_i \hat{X}_{Ii}.
\end{equation}
In the preceding expression, other definitions have been made of $\hat{x}_I \equiv \hat{S}^\dagger (\hat a^\dagger +\hat a) \hat{S}$ and
$\hat{X}_{Ii} \equiv e^{i\Omega_i \hat{b}_i^\dagger \hat{b}_i t} \hat X_i e^{-i\Omega_i \hat{b}_i^\dagger \hat{b}_i t}$,
where $\hat X_i=\kappa_i(\hat b^\dagger_i+\hat b_i).$
Next, by direct integration, we obtain a formal solution to Eq.~(\ref{IntqLiouville}) and resubstitute the outcome into Eq.~(\ref{IntqLiouville}).
Subsequently, we take the trace over the reservoir states to obtain
\begin{eqnarray}
\label{IntSysEq}
\dot{\tilde\rho}(t) =&-&\frac{i}{\hbar} {\rm Tr}_R [\hat{V}_{I}(t),{\tilde\rho}_{tot}(0)] \\
&-& \frac{1}{\hbar^2} \int\limits_0^t  {\rm Tr}_R [\hat{V}_{I}(t),[\hat{V}_{I}(t^\prime),{\tilde\rho}_{tot}(t^\prime)]] dt^\prime, \nonumber
\end{eqnarray}
where ${\tilde\rho}(t)$ is the reduced density operator in the interaction representation, ${\tilde\rho}(t) =\hat{S}^\dagger \hat\rho(t) \hat{S}$.

Other than the contraction of the reservoir degrees of freedom, the obtained Eq.~(\ref{IntSysEq}) is still exact.
We need to furnish some approximations here to obtain a closed equation for ${\tilde\rho}(t)$.
At $t=0$, we suppose that the total density operator is factorized as
\[
\tilde\rho_{tot}(0) = \tilde\rho(0)\otimes\hat\rho_R(0),
\]
where $\hat\rho_R(0)$ is the density operator of the reservoir in equilibrium.
Then, one can prove that the first term on the RHS of Eq.~(\ref{IntSysEq}) vanishes identically.
\[ {\rm Tr}_R [\hat{V}_{I}(t),{\tilde\rho}_{tot}(0)] = [\hat x_I,\hat\rho(0)]{\rm Tr}_R \left(\hat\rho_R (0) \sum \hat{X}_{Ii}\right) \rightarrow 0 \]
Later at $t> 0$, the correlation builds up to the extent that one cannot write the total density operator as a product of the system and reservoir density operators.
Accordingly, we approximate the total density operator to the linear order in the coupling as
\begin{equation*}
{\tilde\rho}_{tot}(t) \cong {\tilde\rho}(t)\otimes\hat\rho_R(0) + {\cal O}(\hat V),
\end{equation*}
which truncates the second term on the RHS of Eq.~(\ref{IntSysEq}) at the second order in the interaction.
Moreover, we neglect the memory effect by replacing the time-dependence of ${\tilde\rho}(t^\prime)$ over the past period $(0,t)$ with its value at present time ${\tilde\rho}(t)$.
Within this standard Born--Markov approximation \cite{Carmichael2002}, we obtain the intended closed equation for the reduced density operator of the system as
\begin{equation}
\label{IntSysEq3}
\dot{\tilde\rho}(t) = - \frac{1}{\hbar^2} \int\limits_0^t  {\rm Tr}_R [\hat{V}_{I}(t),[\hat{V}_{I}(t^\prime),{\tilde\rho}(t)\otimes\hat\rho_R(0)]] dt^\prime.
\end{equation}

Next, we substitute $\hat{V}_{I}$ given in Eq.~(\ref{IntV}) into Eq.~(\ref{IntSysEq3}), and manipulate the commutators to obtain
\begin{widetext}
\begin{eqnarray}
\label{IntSysEq4}
\dot{\tilde\rho}(t) = &-& \frac{1}{\hbar^2}\int \limits_0^t dt^\prime\left\{[\hat{x}_I(t),\hat{x}_I(t^\prime)\tilde{\rho} (t) ] {\rm Tr}_R\bigg( \sum_i \hat{X}_{Ii}(t) \sum_j \hat{X}_{Ij} (t^\prime) \hat\rho_R (0)\bigg) \right.\nonumber\\
&&\left.- [\hat{x}_I(t),\tilde{\rho} (t) \hat{x}_I(t^\prime)] {\rm Tr}_R\bigg( \sum_i \hat{X}_{Ii}(t^\prime) \sum_j \hat{X}_{Ij}(t) \hat\rho_R (0)\bigg) \right\}.
\end{eqnarray}
\end{widetext}
The expressions appearing in the integrand on the RHS of the preceding equation can be further manipulated to produce, for instance,
$
{\rm Tr}_R [\sum_i \hat{X}_{Ii}(t) \sum_j \hat{X}_{Ij} (t^\prime) \hat\rho_R (0)] = \sum_i {\rm Tr}_R [\hat{X}_{Ii} (t-t^\prime) \hat{X}_i \hat\rho_R (0)].
$
The trace in this case can be explicitly evaluated to obtain
\begin{widetext}
\[
{\rm Tr}_R\left( \hat{X}_{Ii} (t-t^\prime) \hat{X}_i \hat\rho_R (0)\right) \rightarrow \kappa_i^2\left\{e^{-i\Omega_i (t-t^\prime)} (\bar{n}(\Omega_i)+1)+e^{i\Omega_i (t-t^\prime)} \bar{n}(\Omega_i)\right\},
\]
\end{widetext}
where $\bar{n}(\Omega_i)$ is the thermal average occupancy of the Debye mode $\Omega_i$, given as
$\bar{n}(\Omega_i) = 1/(e^{\beta\hbar\Omega_i}-1)$, where $\beta=1/(k_BT)$ with $k_B$ being the Boltzmann constant.
Subsequently, in the thermodynamic limit, we replace the summation over the reservoir modes with the continuum identity, $\sum_i {\rm Tr}_R\left( \hat{X}_{Ii} (t-t^\prime) \hat{X}_i \hat\rho_R (0)\right)\equiv  A(t^\prime-t)$, where
\begin{widetext}
\begin{equation}\label{Debye}
A(t^\prime-t)=\int_0^\infty d\Omega g(\Omega) \kappa(\Omega)^2\left\{e^{-i\Omega (t-t^\prime)} (\bar{n}(\Omega)+1)+e^{i\Omega (t-t^\prime)} \bar{n}(\Omega)\right\}, \nonumber
\end{equation}
\end{widetext}
where $g(\Omega)$ is the density of modes limited by the Debye frequency $\Omega_D$,
$g(\Omega) = \frac{V}{2 \pi^2 c_s^3} \Omega^2 \quad{\rm for}\quad  \Omega\le\Omega_D$.
Then, by utilizing the obtained identity $A(t)$, we can convert Eq.~(\ref{IntSysEq4}) into
\begin{widetext}
\begin{eqnarray}
\label{qkinetic}
\dot{\tilde\rho}(t) &=& -\frac{1}{\hbar^2}\int \limits_0^t  dt^\prime\Big\{[\hat{x}_I(t),\hat{x}_I(t^\prime)\tilde{\rho} (t) ] {A}(t-t^\prime)
- [\hat{x}_I(t),\tilde{\rho} (t) \hat{x}_I(t^\prime)] {A}(t^\prime-t)\Big\} .
\end{eqnarray}
\end{widetext}

Before continuing, we consider that the obtained Eq.~(\ref{qkinetic}) should be further discussed.
This equation takes the form of the well-known Bloch--Redfield equation \cite{Blum1981}, except that the dependence on time of a Heisenberg operator $\hat{x}_I(t)= \hat{S}^\dagger(t) \hat{x} \hat{S}(t)$ is given via the unconventional evolution operator $\hat{S}(t)$ defined in Eq.~(\ref{time-devel)}).
Because the generator of the time-translation in $\hat{S}(t)$ contains non-quadratic terms of the creation and annihilation operators, our quantum kinetic equation is not in the Lindblad form (see, for instance, \cite{Andre2012}).

We now consider the matrix representation of the preceding operator equation in the basis of the eigenstates of $\hat H_{rwa}$ defined in Eq.~(\ref{eigen1}).
To this end, we introduce ${\tilde\rho}_{ij}(t) = \langle\phi_i|\tilde{\rho}(t)|\phi_j\rangle$, $\hat{a}^+_{ij} \equiv \left\langle \phi_i\right|\hat{a}^+\left|\phi_j\right\rangle = \hat{a}^*_{ji}$,
and the auxiliary matrix elements
\begin{widetext}
\begin{eqnarray*}\label{matrixeq3}
{B}_{lk}(t) &\equiv& \int_0^t d\tau {A}(\tau) e^{-i \omega_{lk} \tau} \Big( e^{i\omega(t-\tau)}\hat{a}^\dagger_{lk} + e^{-i\omega (t-\tau)}\hat{a}_{lk} \Big),\\
\noindent {C}_{lk}(t) &\equiv& \int_0^t d\tau {A}(-\tau) e^{-i \omega_{lk} \tau} \Big(e^{i\omega (t-\tau)}\hat{a}^\dagger_{lk} + e^{-i\omega(t-\tau)}\hat{a}_{lk} \Big),
\end{eqnarray*}
\end{widetext}
where $\omega_{ij}\equiv ({\tilde E}_i-{\tilde E}_j)/\hbar$.
To proceed further, we have numerically confirmed that for a wide range of physical parameters, the Debye integral $A(\tau)$ defined in Eq.~(\ref{Debye}) decays fast with time.
Accordingly, to a good approximation, we extend the integration limit of the finite time $t$ in the preceding functions to infinity
$\int_0^t d\tau {A}(\tau)\{\cdots\} \approx \int_0^\infty d\tau {A}(\tau)\{\cdots\}$.
In this work, we are mainly interested in the time-evolution of the occupation probability at an arbitrary level $j$.
Accordingly, we consider only the rate of the diagonal term, $i=j$ in  ${\tilde\rho}_{ij}(t)$.
Furthermore, We note that the ensuing terms such as $e^{i(\omega_{jk}\pm2\omega)t}\hat{a}^\dagger_{jl}\hat{a}^\dagger_{lk}$ with level spacing $|\omega_{jk}|\approx 2\omega$ can be dropped because, for this condition, $\hat{a}^\dagger_{jl}\hat{a}^\dagger_{lk}$ are very small, based on numerical observation, for all $l$s.
Furthermore, for those terms $e^{i\omega_{lk}t}{a}^\dagger_{kj}\hat{a}^\dagger_{jl}$ with $|\omega_{jk}|$ not close to $2 \omega$, we make a random phase approximation (RPA) of retaining only the term with $l=k$.
This is because such terms occur only in summations and the terms with $l\neq k$ average out approximately to produce a negligible contribution.
This procedure corresponds to the SVAA exercised in classical Eqs.~(\ref{effeq1}) and (\ref{effeq2}), wherein all terms with fast oscillatory dependence were dropped out.
Consequently, we obtain the coarse-grained density matrix equation, which involves only diagonal matrix elements, as
\begin{widetext}
\begin{eqnarray}\label{diagonal2}
\dot{\tilde{\rho}}_{jj} = - \frac{1}{\hbar^2} && \Big\{ \sum_{l} \tilde{\rho}_{jj}\Big(
 |a_{jl}|^2(D_>(\omega_{lj}+\omega)+D_<(\omega_{jl}-\omega)) + |a_{lj}|^2(D_>(\omega_{lj}-\omega) + D_<(\omega_{jl}+\omega)) \Big)\nonumber\\
&& -\sum_l \tilde{\rho}_{ll} \Big( |a_{jl}|^2(D_>(\omega_{jl}-\omega)+D_<(\omega_{lj}+\omega)) + |a_{lj}|^2(D_>(\omega_{jl}+\omega)+D_<(\omega_{lj}-\omega)) \Big)\Big\},
\end{eqnarray}
\end{widetext}
where we have introduced the auxiliary functions
\begin{eqnarray*}
&&D_>(\omega_{lk}\pm\omega) \equiv \int_0^\infty d\tau e^{- i(\omega_{lk}\pm\omega)\tau} A(\tau), \\
&&D_<(\omega_{lk}\pm\omega)\equiv \int_0^\infty d\tau e^{-i(\omega_{lk}\pm\omega)\tau} A(-\tau).
\end{eqnarray*}
In Eq.~(\ref{diagonal2}), we combine the auxiliary functions and manipulate them to formulate the delta functions specifying the energy exchange of the Josephson particle with photons ($\omega$) and phonons ($\Omega$) as
${\tilde E}_j-{\tilde E}_l = \hbar\omega \pm \hbar\Omega$.
For instance, after some manipulation, one can obtain
\begin{widetext}
\begin{eqnarray*}
D_>(\omega_{lj}+\omega)+D_<(\omega_{jl}-\omega)
= 2\pi\kappa^2 \bigg\{g(-\omega_{lj}-\omega)(\bar{n}(-\omega_{lj}-\omega)+1) + g(\omega_{lj}+\omega)\bar{n}(\omega_{lj}+\omega) \bigg\}.
\end{eqnarray*}
\end{widetext}

Similarly, for another combination, we obtain
\begin{widetext}
\begin{equation*}
\label{eq:delta3}
D_>(\omega_{lj}-\omega)+D_<(\omega_{jl}+\omega) = 2\pi\kappa^2 \bigg\{g(-\omega_{lj}+\omega)(\bar{n}(-\omega_{lj}+\omega)+1) + g(\omega_{lj}  -\omega)\bar{n}(\omega_{lj}-\omega)\bigg\}.
\end{equation*}
\end{widetext}

Finally, by substituting the simplified expressions for $D_>(\omega_{lj}\pm\omega)+D_<(\omega_{jl}\mp\omega)$ into Eq.~(\ref{diagonal2}), we obtain the desired master equation, Eq.~(\ref{Pauli}), describing the dissipative dynamics of the occupation probability $P_j$  of the Josephson particle at an arbitrary RWA level $P_j(t)\equiv {\tilde\rho}_{jj}(t)$.
Note that the diagonal elements of the density operator in the interaction picture are identical to those taken in the quasi-stationary states of the original Hamiltonian.
$\tilde\rho_{jj}= \left\langle \phi_j\right| \tilde\rho \left| \phi_j\right\rangle  = \left\langle \phi_j\right| \hat{S}^\dagger \hat\rho \hat{S} \left| \phi_j\right\rangle=\left\langle \Psi_{ E_j}(t)\right| \hat\rho \left| \Psi_{ E_j}(t)\right\rangle = \rho_{jj}$.


\end{document}